\documentclass[useAMS,usenatbib]{mn2e}

\usepackage{aas_macros}

\usepackage{graphicx}

\title[The environmental dependence of the stellar mass-size relation in STAGES galaxies]{The environmental dependence of the stellar 
mass-size relation in STAGES galaxies}

\author[D.~T.~Maltby et al.]
{David~T.~Maltby$^{1}$\thanks{E-mail: ppxdtm@nottingham.ac.uk},
 Alfonso~Arag{\'o}n-Salamanca$^{1}$, 
 Meghan~E.~Gray$^{1}$, 
 Marco~Barden$^{2}$, 
 \newauthor Boris~H{\"{a}}u{\ss}ler$^{1}$,
 Christian~Wolf$^{\,3}$,
 Chien~Y.~Peng$^{4,5}$,
 Knud~Jahnke$^{6}$, 
 \newauthor Daniel H. McIntosh$^{7}$,
 Asmus B\"{o}hm$^{2}$,
 Eelco van Kampen$^{8}$\\
$^{1}$School of Physics and Astronomy, The University of Nottingham, University Park, Nottingham, NG7 2RD, UK. \\
$^{2}$Institute of Astro- and Particle Physics, University of Innsbruck, Technikerstr. 25/8, A-6020 Innsbruck, Austria. \\
$^{3}$Department of Physics, Denys Wilkinson Building, University of Oxford, Keble Road, Oxford, OX1 3RH, UK. \\
$^{4}$NRC Herzberg Institute of Astrophysics, 5071 West Saanich Road, Victoria, V9E 2E7, Canada. \\
$^{5}$Space Telescope Science Institute, 3700 San Martin Drive, Baltimore, MD 21218, USA. \\
$^{6}$Max-Plank-Institut f{\"{u}}r Astronomie, K{\"{o}}nigstuhl 17, D-69117, Heidelberg, Germany. \\
$^{7}$Department of Physics, 5110 Rockhill Road, University of Missouri-Kansas City, Kansas City, MO 64110, USA. \\
$^{8}$European Southern Observatory, Karl-Schwarzschild-Str. 2, D-85748, Garching bei M{\"{u}}nchen, Germany.}

\begin{document}

\date{Accepted Year Month Date. Received Year Month Date; in original form Year Month Date}

\pagerange{\pageref{firstpage}--\pageref{lastpage}} \pubyear{0000}

\maketitle

\label{firstpage}

\begin{abstract}
We present the stellar mass-size relations for elliptical, lenticular, and spiral galaxies in the field and cluster environments 
using {\em HST}/ACS imaging and data from the Space Telescope A901/2 Galaxy Evolution Survey (STAGES). We use a large sample of 
$\sim1200$ field and cluster galaxies, and a sub-sample of cluster core galaxies, and quantify the significance of any putative 
environmental dependence on the stellar mass-size relation. For elliptical, lenticular, and high-mass (${\rm log}M_*/{\rm M_{\odot}}>10$) 
spiral galaxies we find no evidence to suggest any such environmental dependence, implying that internal drivers are governing their 
size evolution. For intermediate/low-mass spirals (${\rm log}M_*/{\rm M_{\odot}}<10$) we find evidence, significant at the $2\sigma$ 
level, for a possible environmental dependence on galaxy sizes: the mean effective radius $\overline{a}_e$ for lower-mass spirals is 
$\sim15$-$20$ per cent larger in the field than in the cluster. This is due to a population of low-mass large-$a_e$ field spirals that 
are largely absent from the cluster environments. These large-$a_e$ field spirals contain extended stellar discs not present in their 
cluster counterparts. This suggests the fragile extended stellar discs of these spiral galaxies may not survive the environmental 
conditions in the cluster. Our results suggest that internal physical processes are the main drivers governing the size evolution of 
galaxies, with the environment possibly playing a role affecting only the discs of intermediate/low-mass spirals.

\end{abstract}



\section[]{Introduction}

\label{Introduction}

It is now well established that there exist correlations between the properties of galaxies and the environment in which they are 
found. These correlations suggest that local and large scale environment can affect e.g. morphology, colour, and star formation rate 
\citep[e.g.,][]{Dressler:1980,Weinmann_etal:2006}. Theory attempts to account for these environmental effects by, for example, ram 
pressure stripping of the interstellar medium, mergers, and harassment \citep[e.g.,][]{Gunn_Gott:1972,Okamoto_Nagashima:2004,
Moore_etal:1996}. These processes may cause some galaxies to e.g. lose their extended gas haloes \citep[e.g.,][]{Larson_etal:1980}, 
or cause disk truncations \citep{vanderKruit:1979}, and hence affect the physical size of the galaxy.

Recent observations have found that massive galaxies at high redshift ($z>1$) are much more compact than galaxies of the same stellar 
mass in the local universe \citep{Daddi_etal:2005,Trujillo_etal:2006b,Longhetti_etal:2007,Cimatti_etal:2008,vanDokkum_etal:2008}. The 
extreme rarity of these compact high-mass objects in the local universe \citep{Shen_etal:2003,Cimatti_etal:2008,Trujillo_etal:2009} 
implies a strong size evolution in massive galaxies. Various studies have characterized this strong size evolution for massive 
galaxies between $z\sim1.5$ and $z=0$ \citep{Trujillo_etal:2007,Rettura_etal:2008,vanderWel_etal:2008} and out to higher redshifts 
$z\sim3 $ \citep{Trujillo_etal:2006a,Toft_etal:2007,Zirm_etal:2007,Buitrago_etal:2008,Franx_etal:2008,vanDokkum_etal:2008}; the size 
evolution is especially strong for high-mass galaxies ($M_*>10^{10}{\rm M_{\odot}}$) with low star formation rates 
\citep{Franx_etal:2008}. The strength of size evolution also depends on morphology. \cite{Trujillo_etal:2007} find that, for massive 
galaxies at a given stellar mass, disc-like objects were a factor of 2 smaller at $z\sim1.5$ than their counterparts at $z=0$. The 
size evolution is even stronger for spheroid-like galaxies which are a factor of 4 smaller at $z\sim1.5$ compared to analogous 
spheroids in the local universe. The observed size evolution is stronger than, but, in qualitative agreement with hierarchical 
semi-analytical model predictions \citep[e.g.,][]{Khochfar_Silk:2006}.

Systematic uncertainties (e.g. errors in photometric redshifts and mass measurements) could potentially have hampered previous 
studies of size evolution. However, \cite{vanderWel_etal:2008} find, using dynamical masses and spectroscopic redshifts, that these 
systematic effects are too small to account for the observed size evolution.

Several processes have been suggested to account for the observed size evolution of galaxies. One such process is dissipationless or 
`dry' merging \citep[without star formation;][]{Bell_etal:2005,vanDokkum:2005}. Due to the reduced amounts of available gas, `dry' 
mergers are efficient in increasing the physical size of the stellar distribution but inefficient at causing new star formation. 
Another possible process is satellite or smooth envelope accretion \citep{Naab_etal:2007}, where accreted stars from minor/major 
mergers form an envelope around the compact remnant the size of which increases smoothly with decreasing redshift. If this is the 
case we might expect to find the compact core hidden within early-type galaxies in the local universe. As we expect the merger rate 
to depend on environment, both these processes are environment dependent and may cause a growth in size over time. However, mergers 
also cause an increase in the stellar mass and to first order move galaxies roughly parallel to the mass-size relation 
\citep{vanDokkum_etal:2008}.  An alternative environmental independent process has been proposed by \cite{Fan_etal:2008} who argue 
that for massive spheroidal galaxies ($M_*>2\times10^{10}{\rm M_{\odot}}$) the observed size evolution is directly related to the 
rapid mass loss driven by quasar feedback, where the amount of cold gas removed from the central regions of the galaxy can be enough 
to drive a large increase in the galaxy size by a factor $>3$. Lower mass galaxies would also experience a weaker but non-negligible 
size evolution mainly due to the mass loss powered by stellar winds and supernovae explosions, which is in qualitative agreement with 
observations \citep{Franx_etal:2008}.

Hierarchical models of galaxy evolution predict that early-type galaxies of similar stellar mass in the cluster environment are older
than those in the field due to e.g. accelerated galaxy evolution in dense environments \citep{De_Lucia_etal:2004}. If clusters are 
regions of accelerated evolution one should expect an earlier growth in size for early-type cluster galaxies and a difference in 
galaxy sizes between the field and cluster environments. For late-type galaxies we also expect disk truncations 
\citep{vanderKruit:1979} which may also depend on environment on top of this accelerated evolution. 

Both early- and late-type galaxies follow stellar mass-size relations \citep{Shen_etal:2003,Barden_etal:2005, McIntosh_etal:2005} 
with the physical size increasing as a function of stellar mass. Presently, the little work that has been done on the environmental 
dependence of the stellar mass-size relation has been largely limited to small samples of massive ($M_*>10^{10}{\rm M_{\odot}}$), 
early-type galaxies \citep{Cimatti_etal:2008,Rettura_etal:2008}. Recently, \cite{Cimatti_etal:2008} found a possible trend using a 
small sample of $48$ field and cluster, massive early-type galaxies from the $z = 1.237$ RDCS1252.9-2927 cluster and the Great 
Observatories Origins Deep Survey \citep[GOODS;][]{Giavalisco_etal:2004} South Field. They find a hint that for a fixed redshift 
$z\approx1$ massive cluster early-type galaxies are more preferentially located within the $z\sim0$ mass-size relation than their 
counterparts in lower density environments. However, \cite{Rettura_etal:2008} come to a different conclusion and find no 
environmental dependence of the stellar mass-size relation using a small sample of $45$ field and cluster, massive early-type 
galaxies drawn from the same fields and redshift range as \cite{Cimatti_etal:2008}; they conclude that early-type galaxies must 
undergo a similar size evolution in both environments in order to account for the typical sizes of early-type galaxies at lower 
redshifts.

For late-type galaxies, recent work \citep{Guo_etal:2009,Weinmann_etal:2009} has compared the sizes of satellite and central, 
early- and late-type group galaxies from the Sloan Digital Sky Survey \citep[SDSS;][]{York_etal:2000}. At fixed stellar mass, 
satellite galaxies lie in larger groups/clusters (i.e. denser environments) than central galaxies of similar stellar mass. Both 
\cite{Guo_etal:2009} and \cite{Weinmann_etal:2009} find no difference between the radii of satellite and central early-type galaxies, 
however, at low-masses ($M_* < 10^{10.75}{\rm M_{\odot}}$) , late-type satelites have moderately smaller radii than similar mass 
late-type central galaxies. Similar results have also been reported by \cite{Kauffmann_etal:2004} and \cite{vandenBosch_etal:2008}. 

The aim of this paper is to investigate whether the stellar mass-size relation for different Hubble type morphologies is affected 
by the galaxy environment using larger, more statistically viable field and cluster samples. We construct stellar mass-size relations 
for elliptical, lenticular, and spiral galaxies in the environments of the field, cluster, and cluster core.

The structure of this paper is as follows: in \S\ref{data} we give a brief description of the STAGES dataset relevant to this work. 
We describe the determination of the stellar masses in \S\ref{stellar_masses}, detail the method used for the estimation of galaxy 
size in \S\ref{galaxy_sizes}, and outline our sample selection in \S\ref{sample}. In \S\ref{M-S_relations} we present our observed 
stellar mass-size relations for elliptical, lenticular, and spiral galaxies in the field, cluster, and cluster core environments. 
We provide a discussion of our results in \S\ref{Discussion} and finally, draw our conclusions in \S\ref{conclusions}. Throughout 
this paper, we adopt a cosmology of $H_0=70\,{\rm km s^{-1}Mpc^{-1}}$, $\Omega_\Lambda=0.7$, and $\Omega_m=0.3$, and use total Vega 
magnitudes.


\section[]{Description of the data}

\label{data}

This work is entirely based on the STAGES data published by \cite{Gray_etal:2009}. STAGES (Space Telescope A901/2 Galaxy Evolution 
Survey) is a multi-wavelength survey that covers a wide range of galaxy environments. A complex multi-cluster system at $z\sim0.167$ 
has been the subject of $V$-band (F606W) {\em HST}/ACS imaging covering the full $0.5^{\circ}\times0.5^{\circ}$ ($\sim5\times5\,{\rm 
Mpc}$) span of the multi-cluster system. The ACS imaging is complemented by photometric redshifts and observed-/rest-frame SEDs from 
the 17-band COMBO-17 photometric redshift survey \citep{Wolf_etal:2003}. Extensive multi-wavelength observations using {\em Spitzer},
{\em GALEX}, 2dF, {\em XMM-Newton}, and GMRT have also been carried out. \cite{Gray_etal:2009} have performed S{\'e}rsic profile 
fitting on all {\em HST}/ACS images and conducted simulations to quantify the completeness of the survey, all of which are publicly 
available\footnote{http://www.nottingham.ac.uk/astronomy/stages}.

The COMBO-17 observations used in the STAGES master catalogue were obtained with the Wide Field Imager (WFI) at the MPG/ESO
2.2-m-telescope on La Silla, Chile (see \citealt{Wolf_etal:2003} for further details). COMBO-17 used five broad-band filters 
{\em UBVRI} and 12 medium-band filters covering wavelengths from $350$--$930\,{\rm nm}$ to define detailed optical SEDs for objects 
with $R \leq 24$, with $R$ being the total $R$-band magnitude. Generally, photometric redshifts from COMBO-17 are accurate to $1$ 
per cent in $\delta{z}/(1+z)$ at $R < 21$ which has been spectroscopically confirmed. Photo-$z$ quality degrades for progressively 
fainter galaxies reaching accuracies of $2$ per cent for galaxies with $R \sim 22$ and $10$ per cent for galaxies with $R > 24$ 
\citep{Wolf_etal:2004,Wolf_etal:2008}. The galaxy evolution studies to date on the COMBO-17 data that use photo-$z$ defined galaxy 
samples all restrict themselves to galaxies that are brighter than $R = 24$ to ensure only reliable redshifts are used.

The STAGES morphological catalogue (Gray et al. in prep.) contains 5090 galaxies in STAGES with reliable Hubble type morphologies. All 
galaxies with $R<23.5$ and $z_{\rm phot}<0.4$ were visually classified by seven members of the STAGES team into the Hubble types (E, S0, 
Sa, Sb, Sc, Sd, Irr) and their intermediate classes. S0s were defined to be disc galaxies with a visible bulge but no spiral arms. 
Weighted average estimates of the Hubble types, ignoring bars and degrees of asymmetry were generated. In this paper, we only consider 
the elliptical, lenticular, and spiral classes. The spiral subclasses are merged and we only consider the spiral class as a whole.


\subsection[]{Determination of Stellar Masses}

\label{stellar_masses}

The stellar masses listed in the STAGES catalogue and used in this work were originally estimated by \cite{Borch_etal:2006} for 
galaxies in COMBO-17. These estimates were derived from SED-fitting of the 17-band photometry using a template library derived from 
PEGASE \citep{Fioc_Rocca-Volmerange:1997} population synthesis models and a \cite{Kroupa_etal:1993} stellar IMF. Random errors 
in stellar mass are estimated to be $\sim0.1$ dex on a galaxy-galaxy basis in the majority of cases. Systematic errors in stellar 
mass (for the given population synthesis model and IMF) were argued to be at the $0.1$ dex level for galaxies without ongoing or 
recent major starbursts, however, for galaxies with strong starbursts the stellar mass could be overestimated by up to $\sim0.5$ dex, 
see \cite{Borch_etal:2006} for further details. 

The masses of spiral galaxies are reliable for ${\rm log} M_*/{\rm M_{\odot}}<11$ and the masses of E/S0 galaxies are reliable at all 
masses. However, due to aperture effects on the SEDs, the stellar masses of spiral galaxies are unreliable at ${\rm log}M_*/{\rm 
M_{\odot}}>11$, see \cite{Wolf_etal:2009} for further details. For this reason we limit ourselves to the mass range ${\rm log}M_*/{\rm 
M_{\odot}}=[9,11]$ for spiral galaxies.


\subsection[]{Galaxy Size Determination}

\label{galaxy_sizes}

The galaxy sizes used in this paper are the effective radius along the semi-major axis $a_e$ of the 2D surface brightness distribution.
Historically, work in this field has used the circularized effective radius $r_e = a_e\sqrt{q}$, where $q$ is the axis ratio of the 
galaxy (ratio of semi-minor over semi-major effective radius), as the estimate of galaxy size. However, the use of circularized 
quantities has no effect on the significance of the results of this work or our overall conclusions. Therefore, we choose to use the
more physically meaningful $a_e$ as our measurement of galaxy size. 

\cite{Gray_etal:2009} use the data pipeline GALAPAGOS (Galaxy Analysis over Large Areas: Parameter Assessment by GALFITting Objects 
from SExtractor; Barden et al., in prep) to perform the extraction and S{\'e}rsic model fitting of source galaxies from the 
{\em HST}/ACS $V$-band imaging. This data pipeline uses the GALFIT code \citep{Peng_etal:2002} to fit \cite{Sersic:1968} radial 
surface brightness models to each ACS image.

GALFIT is a two-dimensional fitting algorithm that determines a best-fit model for the observed galaxy surface brightness distribution. 
Two-dimensional \cite{Sersic:1968} $r^{1/n}$ models are convolved with the PSF of the original ACS (F606W) images and compared to the 
original surface brightness distribution. The best-fit model is obtained by minimising the $\chi^2$ of the fit using a 
Levenberg-Marquardt algorithm. GALFIT determines several structural parameters for the galaxy including the effective radius along the 
semi-major axis $a_e$ (in image pixels), axis ratio $q$, and S{\'e}rsic index $n$. The S{\'e}rsic index measures the 
concentration of the surface brightness profile and is used to eliminate GALFIT model fits with potentially unreliable structural 
parameters (i.e. those with S{\'e}rsic indices $n<0.20001$ and $n\geq6$).

We use simulations of the STAGES dataset in order to determine the reliability of the GALFIT structural parameters. 
\cite{Gray_etal:2009} simulated over 10 million galaxy images with a range of properties analogous to the real STAGES data and 
subjected the dataset to the same data pipeline. Using a similar approach to that described by \citep{Haussler_etal:2007}, we determine 
the error in our GALFIT structural parameters by comparing the input and output structural parameters for a simulated sample selected 
by output magnitude and S{\'e}rsic index to match that of our galaxy samples. In this work, we use galaxy samples of 
different morphologies, environments and stellar mass ranges. For each sample we obtain the $R$-magnitude range and use it to select 
an analogous sample from the simulations. The mean error in the GALFIT semi-major axis effective radius as determined from simulations 
(${\rm size_{GALFIT}\;error} = |a_{e({\rm sim})}-a_{e({\rm GALFIT})}|/a_{e({\rm sim})}$) was found to increase slightly with decreasing
stellar mass and be $<10$ per cent in all cases, see Table~\ref{tbl0}.

\begin{table*}
\begin{minipage}{175mm}
\centering
\caption{\label{tbl0}{$R$-magnitude range, mean error in GALFIT galaxy size as determined by simulations (Size$\rm_{GALFIT}$), and total error in 
galaxy size for different morphologies, environments and stellar mass ranges used in this work.}}
\begin{tabular}{lccccccc}
\hline
\hline
{Elliptical Galaxies (E)}            & {}        &{}                 &{}        &{}                         &{}&{}         &{}                           \\
\hline
{${\rm log}M_*/{\rm M_{\odot}}$ range}&\multicolumn{4}{c}{Field}                                             &{}&\multicolumn{2}{c}{Cluster}              \\
\cline{2-5} 
\cline{7-8}
{}                                   &{$R$}      &{Size$\rm_{GALFIT}$}&{Distance}&{Total Size}               &{}&{$R$}     &{Size$\rm_{GALFIT}$:Total Size}\\
{}                                   &{range}    &{error}            &{error}   &{error}                    &{}&{range}    &{error}                      \\
\hline
{[9, 9.5]}	                     &$19.3-23.3$&$8\%$              &$19\%$    &$20\%\;(\pm0.3\,{\rm kpc})$&{}&$19.9-22.7$&$6\%\;(\pm0.1\,{\rm kpc})$   \\
{[9.5, 10]}		             &$20.3-22.0$&$5\%$              &$15\%$    &$15\%\;(\pm0.3\,{\rm kpc})$&{}&$19.6-22.3$&$6\%\;(\pm0.1\,{\rm kpc})$   \\
{[10, 11.5]}		             &$16.8-21.1$&$5\%$              &$5\%$     &$7\%\; (\pm0.3\,{\rm kpc})$&{}&$16.3-19.9$&$5\%\;(\pm0.2\,{\rm kpc})$   \\
\hline
\hline
{Lenticular Galaxies (S0)}           &{}         &{}                 &{}        &{}                         &{}&{}         &{}                           \\
\hline
{${\rm log}M_*/{\rm M_{\odot}}$ range}&\multicolumn{4}{c}{Field}                                             &{}&\multicolumn{2}{c}{Cluster}              \\
\cline{2-5} 
\cline{7-8}
{}                                   &{$R$}      &{Size$\rm_{GALFIT}$}&{Distance}&{Total Size}               &{}&{$R$}     &{Size$\rm_{GALFIT}$:Total Size}\\
{}                                   &{range}    &{error}            &{error}   &{error}                    &{}&{range}    &{error}                      \\
\hline
{[9, 9.5]}		             &$18.8-22.6$&$6\%$              &$20\%$    &$20\%\;(\pm0.4\,{\rm kpc})$&{}&$20.6-22.1$&$6\%\;(\pm0.1\,{\rm kpc})$   \\
{[9.5, 10]}	                     &$18.8-21.9$&$5\%$              &$14\%$    &$14\%\;(\pm0.4\,{\rm kpc})$&{}&$19.7-21.0$&$5\%\;(\pm0.1\,{\rm kpc})$   \\
{[10, 11]}	                     &$18.1-21.0$&$5\%$              &$5\%$     &$7\%\; (\pm0.2\,{\rm kpc})$&{}&$17.6-19.9$&$5\%\;(\pm0.2\,{\rm kpc})$   \\
\hline
\hline
{Spiral Galaxies (Sp)}               &{}         &{}                 &{}        &{}                         &{}&{}         &{}                           \\
\hline
{${\rm log}M_*/{\rm M_{\odot}}$ range}&\multicolumn{4}{c}{Field}                                             &{}&\multicolumn{2}{c}{Cluster}              \\
\cline{2-5} 
\cline{7-8}
{}                                   &{$R$}      &{Size$\rm_{GALFIT}$}&{Distance}&{Total Size}               &{}&{$R$}     &{Size$\rm_{GALFIT}$:Total Size}\\
{}                                   &{range}    &{error}            &{error}   &{error}                    &{}&{range}    &{error}                      \\
\hline
{[9, 9.5]}	                     &$18.7-23.3$&$8\%$              &$15\%$    &$17\%\;(\pm0.5\,{\rm kpc})$&{}&$19.6-22.2$&$6\%\;(\pm0.2\,{\rm kpc})$   \\
{[9.5, 10]}		             &$17.9-22.6$&$6\%$              &$10\%$    &$11\%\;(\pm0.4\,{\rm kpc})$&{}&$18.6-21.0$&$5\%\;(\pm0.2\,{\rm kpc})$   \\
{[10, 11]}		             &$15.5-21.1$&$5\%$              &$8\%$     &$9\%\; (\pm0.4\,{\rm kpc})$&{}&$16.9-20.1$&$5\%\;(\pm0.3\,{\rm kpc})$   \\
\hline
\hline
\end{tabular}
\end{minipage}
\end{table*}

The semi-major axis effective radii $a_e$ were transformed into intrinsic linear scales using a cosmology of $H_0=70\,{\rm 
km s^{-1}Mpc^{-1}}$, $\Omega_\Lambda=0.7$, and $\Omega_m=0.3$. We use the fixed cluster redshift ($z = 0.167$) to determine the sizes of 
our cluster galaxies and the original COMBO-17 redshift estimate for our field galaxies. Therefore, the photo-$z$ errors only propagate 
into the physical sizes of our field galaxies and not our cluster galaxies. The STAGES multi-cluster system is at a distance of 
$\sim600\,{\rm Mpc}$ and has a maximum depth of $20\,{\rm Mpc}$ along the line-of-sight. Consequently, the $a_e$ of cluster galaxies 
are the same ($\Delta{a_e} < 10^{-15}\,{\rm kpc}$) regardless of whether the COMBO-17 redshift is used or the redshift is fixed at 
$z=0.167$. Therefore, fixing the redshift of the cluster sample has no effect on the $a_e$ distribution of the cluster galaxies but 
eliminates the main source of $a_e$ uncertainty (photo-$z$ errors). The mean error in $a_e$ associated with the photo-$z$ error 
(i.e. error in distance to galaxy) for our field galaxies was calculated for each morphology and stellar mass range. The
distance error is larger for lower stellar masses, see Table~\ref{tbl0}, and is $<20$ per cent in all cases. We return to the effect 
of this error in \S\ref{photo-z_errors}. 

The total mean error in our galaxy sizes is $<20$ per cent for our field sample and $<10$ per cent for our cluster sample for all 
stellar mass ranges used in this work.


\subsection[]{Sample Selection}

\label{sample}

\cite{Gray_etal:2009} suggest a cluster sample for STAGES defined solely from photometric redshifts. The photo-$z$ distribution 
of cluster galaxies was assumed to follow a Gaussian, while the field distribution was assumed to be consistent with the average 
galaxy counts $N(z, R)$ outside the cluster and to vary smoothly with redshift and magnitude. The cluster sample is defined by a 
redshift interval $z_{\rm phot}=[0.17-\Delta{z},0.17+\Delta{z}]$, where the half-width $\Delta{z}$ was allowed to vary with 
$R$-magnitude. A narrow redshift range is adopted for bright $R$-magnitudes due to the high precision of COMBO-17 photometric 
redshifts, but the interval increases in width towards fainter $R$-magnitudes to accommodate for the increase in photo-$z$ error. The 
completeness and contamination of the cluster sample at all magnitude points was calculated using the counts of the smooth 
models \citep[see Fig. 14 from][]{Gray_etal:2009} and the photo-$z$ half-width compromised so the completeness was $>90$ per 
cent at any magnitude. Contamination is defined to be the fraction of field galaxies in the cluster sample at a given magnitude (not 
below). The half-width as a function of magnitude $R$ is

\begin{equation}
\Delta{z}(R) = \sqrt{0.015^2 + 0.0096525^2(1 + 10^{0.6(R_{\rm tot} - 20.5)})}.
\end{equation}

This equation defines a photo-$z$ half-width that is limited to 0.015 at bright $R$-magnitudes but increases as a constant multiple 
of the estimated photo-$z$ error at the faint end. The completeness of this selection converges to nearly $100$ per cent for bright 
galaxies; see \cite{Gray_etal:2009} for further details. The catalogue published by \cite{Gray_etal:2009} contains a number of flags 
({\em combo\_flag, phot\_flag, stages\_flag}) that allows the selection of various galaxy samples. 

\begin{figure}
\includegraphics[width=0.45\textwidth]{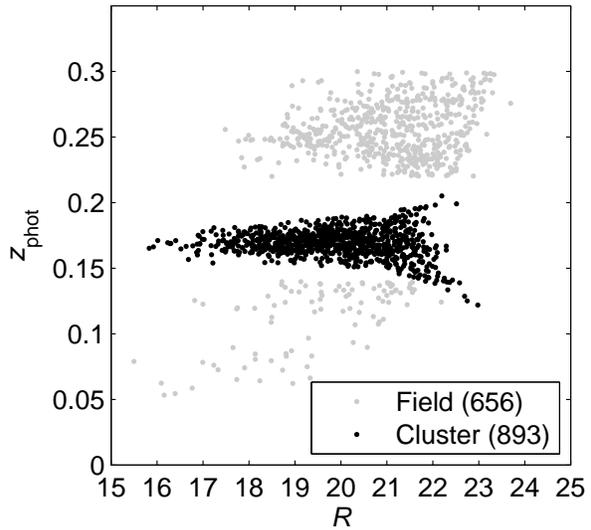}
\caption{\label{cluster_field} The photometric redshift estimate vs. total $R$-band magnitude for the cluster ({\em black points}) and 
field ({\em grey points}) samples, showing there is no overlap between the two samples as might be inferred from their redshift ranges. 
The cluster sample reaches down to $R\sim23$ and the field sample reaches down to $R\sim23.5$. Sample sizes are shown in the legend.}
\end{figure}

For our cluster sample we use the above cluster definition ({\em combo\_flag} $\geq 4$), we also only use galaxies with reliable 
photometry (i.e. those with {\em phot\_flag} $<8$), and those defined as extended {\em HST} sources in STAGES ({\em stages\_flag} 
$\geq 3$). We also limit our sample by stellar mass, cutting at ${\rm log} M_*/{\rm M_{\odot}}>9$. This cluster sample contains 893 
galaxies which reach to $R = 23$ and has a photo-$z$ range at the low-mass end of $z_{\rm phot}=[0.122, 0.205]$, see 
Fig.~\ref{cluster_field}.

For our field sample we use COMBO-17 defined galaxies ({\em combo\_flag} $\geq 3$) and apply a redshift selection that avoids the 
cluster. We include a lower redshift interval at $z = [0.05, 0.14]$ and an upper redshift interval at $z = [0.22, 0.30]$, based on a 
similar sample selection used by \cite{Wolf_etal:2009}. We also only use galaxies with reliable photometry (i.e. those with {\em 
phot\_flag} $<8$), and those defined as extended {\em HST} sources in STAGES ({\em stages\_flag} $\geq 3$). We also limit our sample 
by stellar mass, cutting at ${\rm log}M_*/{\rm M_{\odot}}>9$. This field sample contains 656 galaxies which reach to $R = 23.5$, 
see Fig.~\ref{cluster_field}.

The catalogue published by \cite{Gray_etal:2009} contains two sets of derived values for properties such as magnitude and stellar mass, 
one based on the photo-$z$ estimate and another assuming the galaxy is fixed at the cluster redshift of $z=0.167$. This prevents the 
propagation of photo-$z$ errors into physical values. Here we use the fixed redshift set of values for the cluster sample, but the 
original estimates for our field comparison sample.

Fig.~\ref{cluster_field} shows the redshift-magnitude diagram for our field and cluster samples and shows that there is no overlap 
between the field and cluster samples near $z \simeq 0.13$ as might be inferred from their photo-$z$ ranges in the above discussion, 
see \cite{Wolf_etal:2009} for an explanation. The upper redshift interval for the field contains much more volume and hence more 
galaxies than the lower redshift interval.

\begin{figure}
\includegraphics[width=0.47\textwidth]{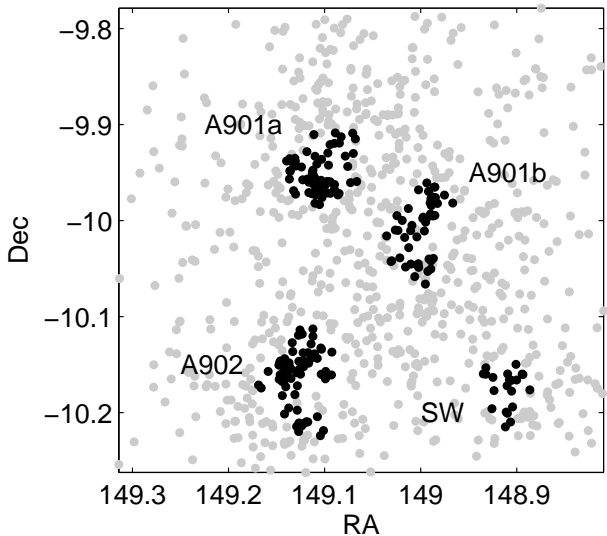}
\caption{\label{core_map} A sky map of the STAGES region showing our mass-selected cluster sample galaxies (${\rm log} M_*/{\rm 
M_{\odot}}>9$). Galaxies located where the aperture stellar mass surface density ${\rm log}\Sigma_{\rm 300 kpc}^{M}(>10^9{\rm M_{\odot}})
>12.5$ ({\em black points}) are designated to lie within the four cores of the multi-cluster system (A901a, A901b, A902, SW group).}
\end{figure}

In this paper, we also consider a comparison of the stellar mass-size relation between the extreme environments of the field and 
cluster core. Any environmental effects on galaxy size will be more apparent in this comparison. \cite{Wolf_etal:2009} create a 
sample of cluster core galaxies in STAGES using the stellar mass surface density of cluster member galaxies. A cluster sample
with ${\rm log}M_*/{\rm M_{\odot}}>9$ was used to measure the stellar mass surface density inside a fixed aperture with radius $r = 
300\,{\rm kpc}$ at the redshift of the multi-cluster system ($z = 0.167$). The aperture stellar mass density $\Sigma_r^M$ in units 
of ${\rm M_{\odot}}\,{\rm Mpc^{-2}}$ was then used to define a cluster core sample. Galaxies where ${\rm log} \Sigma_{\rm 300\,kpc}^{M}
(>10^9{\rm M_{\odot}})>12.5$ were designated to lie within the cluster cores of the STAGES multi-cluster system. We apply the same 
selection as \cite{Wolf_etal:2009} to our cluster sample and obtain a cluster sub-sample of 203 cluster core galaxies. 
Fig.~\ref{core_map} shows the location of our cluster and cluster core sample galaxies in the STAGES region and illustrates the 
extraction of cluster core galaxies from the cluster sample.

\subsubsection[]{Sample Completeness and Morphologies}

\label{sample_completeness}

\begin{figure}
\centering
\includegraphics[width=0.45\textwidth]{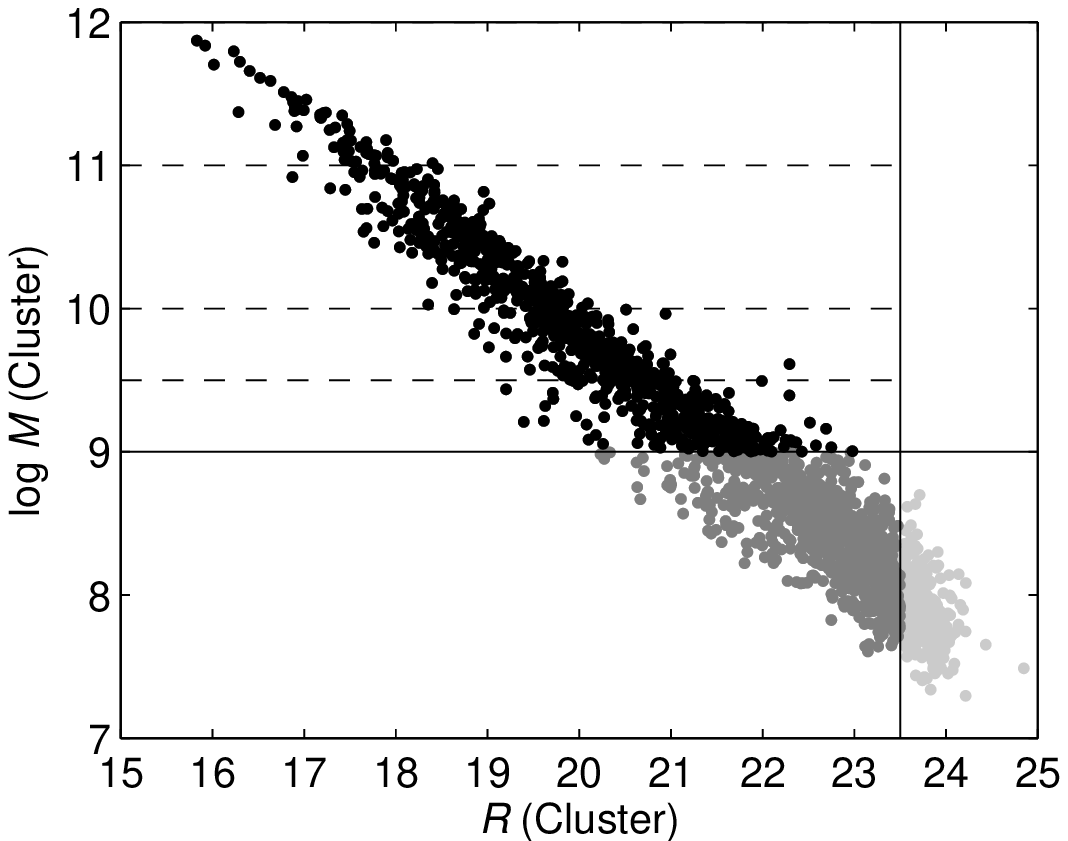}
\includegraphics[width=0.45\textwidth]{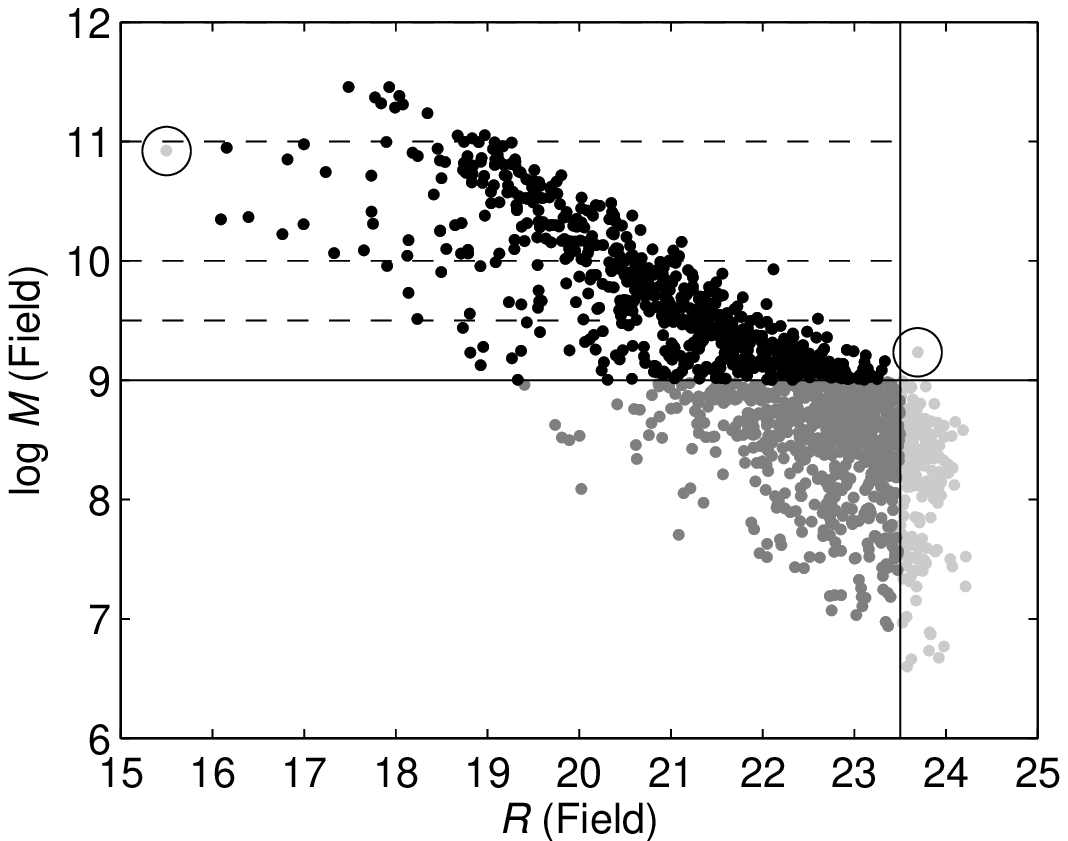}
\caption{\label{M-R_relations} {\em Top panel:} Mass-magnitude diagram for the cluster sample without the stellar mass cut. {\em 
Bottom panel:} Mass-magnitude diagram for the field sample without the stellar mass cut. In both cases, {\em Light grey points:} 
sample galaxies not in the STAGES morphological catalogue. {\em Dark grey points:} sample galaxies in the STAGES morphological 
catalogue and with stellar mass ${\rm log}M_*/{\rm M_{\odot}}<9$. {\em Black points:} sample galaxies in the STAGES morphological 
catalogue and with stellar mass ${\rm log}M_*/{\rm M_{\odot}}>9$. Our mass-selected cluster and field samples are sample galaxies with 
${\rm log}M>9$ and $R<23.5$, where $M=M_*/{\rm M_{\odot}}$. In the field mass-magnitude diagram, the two galaxies highlighted with a 
circle are those in our mass-selected sample without visual morphologies. For both the cluster and field samples, selecting only 
galaxies with visual morphologies essentially introduces no further incompleteness. The dashed lines show the limits of the stellar 
mass bins.}
\end{figure}

We require visual morphologies for our field and cluster samples from the STAGES morphological catalogue. All galaxies with
$z_{\rm phot}<0.4$ (as is the case in this work) and $R<23.5$ have been visually classified into Hubble type morphologies (Gray et al., 
in prep). In order to assess any selection effect or incompleteness introduced by only using galaxies with visual morphologies, 
we plot the mass-magnitude relation for our field and cluster samples without the stellar mass cut, see Fig.~\ref{M-R_relations}. All 
galaxies in our mass-selected cluster sample (${\rm log}M_*/{\rm M_{\odot}}>9$) have $R<23.5$ and have visual morphologies in the 
STAGES morphological catalogue. Therefore, we introduce no further incompleteness to our cluster sample by selecting only galaxies 
that have visual morphologies.

For our field sample, apart from two exceptions all galaxies with (${\rm log}M_*/{\rm M_{\odot}}>9$) have $R<23.5$ and have visual 
morphologies in the STAGES morphological catalogue. Therefore, essentially no further incompleteness is introduced to our field 
sample by selecting only galaxies that have visual morphologies. One bright high-mass galaxy, (object 35364) in our mass-selected 
sample was not contained in the original morphological catalogue due to anomalous photometric flags, but upon subsequent visual 
inspection this galaxy was classified as a spiral. For completeness this spiral was retained in our field sample. A faint ($R>23.5$) 
galaxy (object 5622) in our mass-selected sample is unclassified due to the galaxy being too faint for classification. 

STAGES is $>90$ per cent complete for $R<23.5$ \citep{Gray_etal:2009} as is the case for all galaxies with visual morphologies in the 
morphological catalogue. However, \cite{Wolf_etal:2009} estimate that at masses below ${\rm log}M_*/{\rm M_{\odot}} < 9.5$, the field 
sample could have an addiational $20$ per cent incompleteness based on previous COMBO-17 experience. Therefore, our field sample (all 
with $R<23.5$) is essentially $>90$ per cent complete for ${\rm log}M_*/{\rm M_{\odot}} > 9.5$ and $>70$ per cent complete for ${\rm 
log}M_*/{\rm M_{\odot}} = [9,9.5]$.

In the analysis of this work we consider mass-selected sub-samples of the field and cluster samples, and cluster core sub-sample with 
${\rm log}M_*/{\rm M_{\odot}}$ in the ranges $[9,9.5]$, $[9.5,10]$ and $[10,11]$ or $[10,11.5]$. We expect the completeness of the 
cluster samples to be $>90$ per cent and the contamination of the cluster samples by the field to be $<25$ per cent in all 
cases, based on the $R$-magnitude the respective sub-sample reaches down to \citep{Gray_etal:2009}, see Fig.~\ref{M-R_relations}. 
Table~\ref{tbl1} has detailed numbers per mass bin.

\begin{table}
\begin{minipage}{80mm}
\centering
\caption{\label{tbl1}{Properties of the cluster galaxy sample}}
\begin{tabular}{lcccc}
\hline
\hline
{${\rm log}M_*/{\rm M_{\odot}}$ range}  & {[9, 9.5]} 	& {[9,5, 10]} 	& {[10, 11]} 	& {[11, 12]}             \\
\hline
completeness 		& $>90\%$ 	& $>95\%$ 	& $>95\%$ 	& $>95\%$ 	         \\
contamination 		& $<25\%$ 	& $<15\%$ 	& $<10\%$ 	& $<5\%$	         \\
$R$ mean        	& 21.22 	& 20.11 	& 18.83 	& 17.21 	         \\
$z_{\rm phot,min}$ 	& 0.122 	& 0.151 	& 0.154 	& 0.154 	         \\
$z_{\rm phot,max}$ 	& 0.205 	& 0.190 	& 0.187 	& 0.181 	         \\
$N_{\rm gal}$ 		& 299 		& 236 		& 302 		& 56 		         \\
$N_{\rm gal}$ (Field) 	& 308 		& 163 		& 172 		& 11	 	         \\
\hline
\end{tabular}
\end{minipage}
\end{table}

We separate the field and cluster samples, and the cluster core sub-sample into Hubble type morphologies using the STAGES 
morphological catalogue (Gray et al., in prep.) and obtain a sample of elliptical, lenticular (S0), and spiral galaxies in each 
environment. We obtain 477 field galaxies (100 E, 60 S0, 317 Sp) and 791 cluster galaxies (192 E, 216 S0, 383 Sp), 190 of which are 
cluster core galaxies (64 E, 67 S0, 59 Sp). Irregular galaxies and other objects are not considered in this paper.

\subsubsection[]{Reliability of Structural Parameters}

\label{error}

We remove all high-mass (${\rm log}M^*/{\rm M_{\odot}}>11$) spirals from our samples (14 cluster and 1 field) due to their unreliable 
stellar masses \citep{Wolf_etal:2009}, see section~\ref{stellar_masses}. We also remove poor GALFIT \citep{Peng_etal:2002} model fits 
(i.e. those with S{\'e}rsic indices $n<0.20001$ and $n\geq6$) due to the potential unreliable nature of their structural parameters 
(see section~\ref{galaxy_sizes}). This removes 22 galaxies ($\sim4.6$ per cent) from the field sample (11 E, 1 S0 and 10 Sp) and 43 
galaxies ($\sim5.4$ per cent) from the cluster sample (20 E, 13 S0 and 10 Sp), 8 galaxies of which are from the cluster core 
sub-sample (6 E, 1 S0 and 1 Sp), and removes outliers (galaxies with incorrect $a_e$) from our stellar mass-size relations.
Some authors consider high sersic indices intrinsic for galaxies \citep[e.g.][]{Graham_etal:1996}, however, this S{\'e}rsic cut only 
affects the significance of our results for high-mass ellipticals where it removes some of the Brightest Cluster Galaxies (BCGs). 
Otherwise, this S{\'e}rsic cut has no effect on the overall significance of the results of this 
work or our conclusions. The BCGs of the STAGES multi-cluster system are poorly fitted by GALFIT due to large amounts of inter-cluster 
light. However, 2 are removed by the S{\'ersic} data-cut and the remainder (3 BCGs) are all high-mass (${\rm log}M_*/{\rm M_{\odot}}>
11.5$) ellipticals which are not considered in the comparisons of our mass-size relations (see section~\ref{M-S_relations}). This 
leaves a final sample of 455 field galaxies (89 E, 59 S0, 307 Sp) and 734 cluster galaxies (172 E, 203 S0, 359 Sp), 177 of which are 
cluster core galaxies (58 E, 66 S0, 53 Sp).


\section[]{The Observed Stellar Mass-Size Relations}

\label{M-S_relations}

\begin{figure*}
\includegraphics[width=1\textwidth]{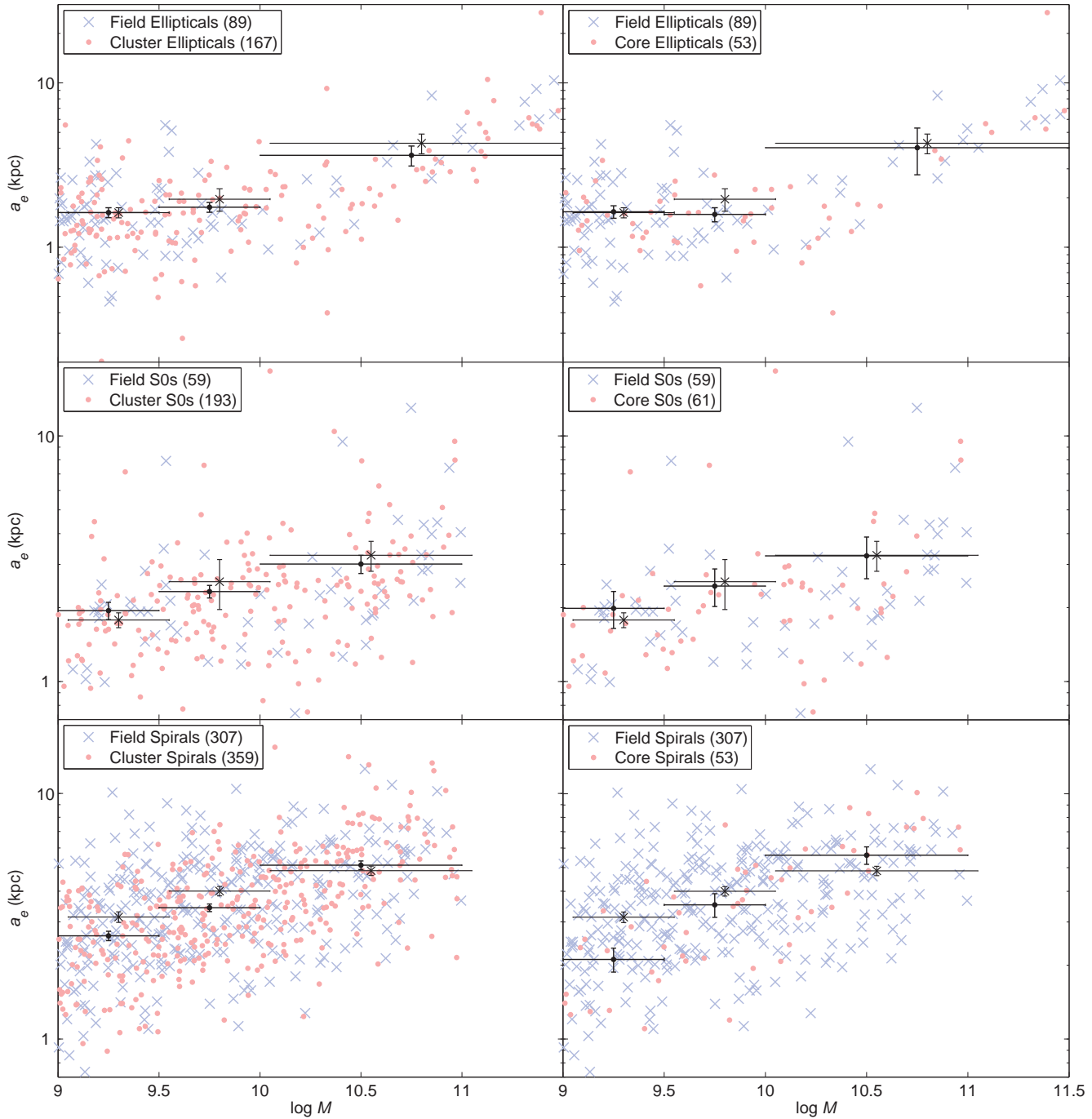}
\centering
\caption{\label{M-S} The stellar mass-size relations for, {\em top row:} elliptical galaxies, {\em middle row:} lenticular galaxies 
(S0), and {\em bottom row:} spiral galaxies. {\em Left panels:} A comparison of the stellar mass-size relation in the environments of 
the field ({\em blue crosses}), and the cluster ({\em red points}). {\em Right panels:} A comparison of the stellar mass-size relation 
in the environments of the field ({\em blue crosses}) and the cluster {\em core} ({\em red points}). For each mass bin (represented by 
the $x$-axis error bars), we overplot on the observed distributions the mean semi-major axis effective radius $\overline{a}_e$ in the 
field and the cluster/core. The $\overline{a}_e$ error bars are the uncertainty ($1\sigma$) in the mean. For display purposes, the 
field $\overline{a}_e$ values are displaced in stellar mass by $+0.05$ dex. There are no spiral or lenticular galaxies with ${\rm log}
M>11$ as these have explicitly been removed from our stellar mass-size relations, see section~\ref{M-S_relations}. We observe no 
significant difference between the stellar mass-size relations in each environment for each morphological type, except for 
intermediate/low-mass spirals. For stellar masses ${\rm log} M<10$ we find that the cluster spirals have values of $\overline{a}_e$ 
significantly smaller than the $\overline{a}_e$ for field spirals of the same mass. We also find that low-mass spirals (${\rm log} 
M<9.5$) in the core have $\overline{a}_e$ significantly smaller than similar galaxies in the field. The mean relative size error is 
$<20$ per cent and uncertainties in the stellar mass are $\sim0.1$ dex. Respective sample sizes are shown in the legends and $M=M_*/
{\rm M_{\odot}}$.} 
\end{figure*}

In this section, we discuss the stellar mass-size relations for our sample of field and cluster galaxies, and our cluster sub-sample 
of cluster core galaxies. For each Hubble type we compare the stellar mass-size relations between the field and cluster, and the field 
and cluster core. These comparisons are presented in Fig.~\ref{M-S} for elliptical, lenticular, and spiral galaxies.

For each Hubble type, the field and cluster samples, and cluster core sub-sample are split into three stellar mass (${\rm log} 
M_*/{\rm M_{\odot}}$) bins. The two low-mass bins, $[9, 9.5]$ and $[9.5, 10]$, are the same for all Hubble types but the high-mass bin 
varies. We use a high-mass bin of $[10, 11]$ for spiral and lenticular galaxies but extended it to $[10, 11.5]$ for elliptical 
galaxies. For spiral galaxies, we do not consider ${\rm log} M_*/{\rm M_{\odot}}>11$ due to unreliable stellar masses 
\citep{Wolf_etal:2009}, see section~\ref{stellar_masses}. For lenticular galaxies, we do not consider ${\rm log} M_*/{\rm M_{\odot}}>11$ 
due to there being no field lenticulars with $M_* > 10^{11}{\rm M_{\odot}}$ to compare to cluster lenticulars. We remove 10 cluster S0s, 
5 of which are from the cluster core, with $M_*>10^{11}{\rm M_{\odot}}$ from our stellar mass-size relations. For elliptical galaxies, 
we do not consider ${\rm log} M_*/{\rm M_{\odot}}>11.5$ due to there being no field ellipticals with $M_*>10^{11.5}{\rm M_{\odot}}$ to 
compare to cluster ellipticals. We remove 5 cluster ellipticals (including 3 BCGs), all from the cluster core, with $M_*>10^{11.5}
{\rm M_{\odot}}$ from our stellar mass-size relations.

For each Hubble type and environment, we calculate the mean semi-major axis effective radius $\overline{a}_e$ with associated 
$1\sigma$ uncertainty in the mean for each stellar mass bin, see Table~\ref{results}. The values of $\overline{a}_e$ with $1\sigma$ 
errorbars are overplotted on our stellar mass-size relations, see Fig.~\ref{M-S}. For elliptical and lenticular galaxies we find no 
significant difference between the values of $\overline{a}_e$ for the field and cluster/core samples in all mass bins. For high-mass 
(${\rm log} M_*/{\rm M_{\odot}}>10$) spiral galaxies, we find no significant difference between the value of $\overline{a}_e$ for the 
field and cluster samples, however, we find core spirals have $\overline{a}_e$ slightly larger than the field spirals. For lower 
stellar masses (${\rm log} M_*/{\rm M_{\odot}}<10$) we find that the cluster spirals have values of $\overline{a}_e$ significantly 
smaller than the $\overline{a}_e$ for field spirals of the same mass. We also find that low-mass spirals (${\rm log} M_*/{\rm 
M_{\odot}}<9.5$) in the core have $\overline{a}_e$ significantly smaller than similar galaxies in the field, see Table~\ref{results}. 
Using the median-$a_e$ instead of the mean-$a_e$ only removes the difference in average size between the low-mass (${\rm log} M_*/{\rm 
M_{\odot}}<9.5$) spirals of the cluster and the field. Otherwise, these results are robust to the use of median-$a_e$.

\begin{figure*}
\includegraphics[width=0.3\textwidth]{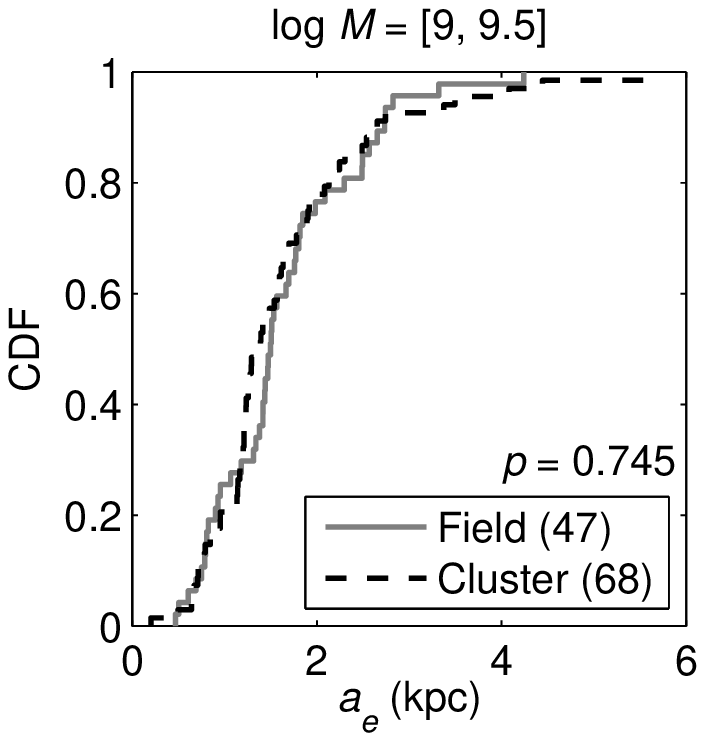}
\includegraphics[width=0.3\textwidth]{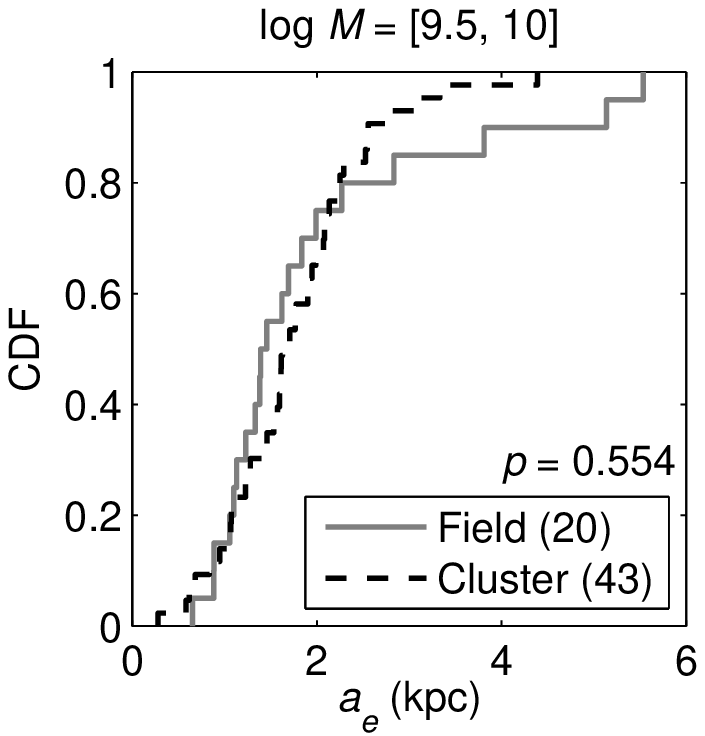}      
\includegraphics[width=0.3\textwidth]{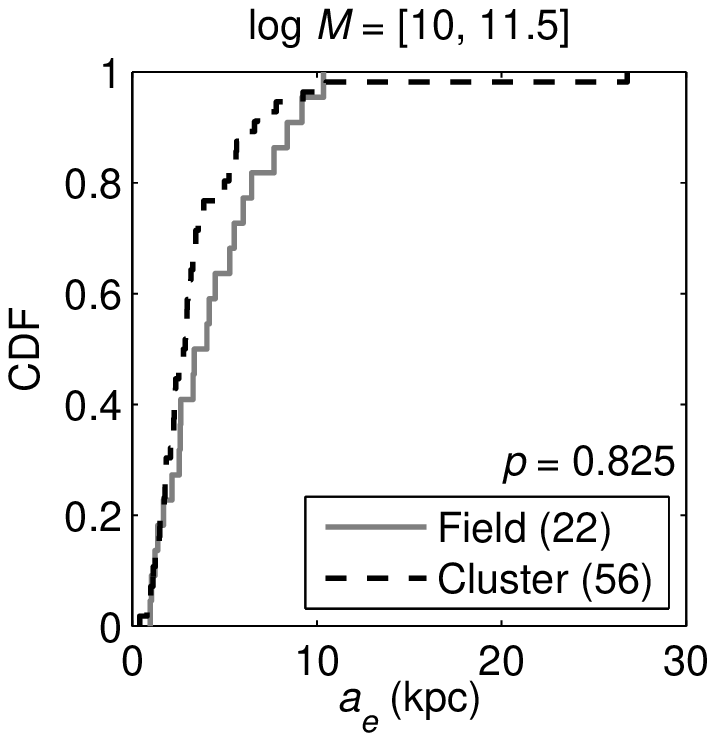}      \\
\includegraphics[width=0.3\textwidth]{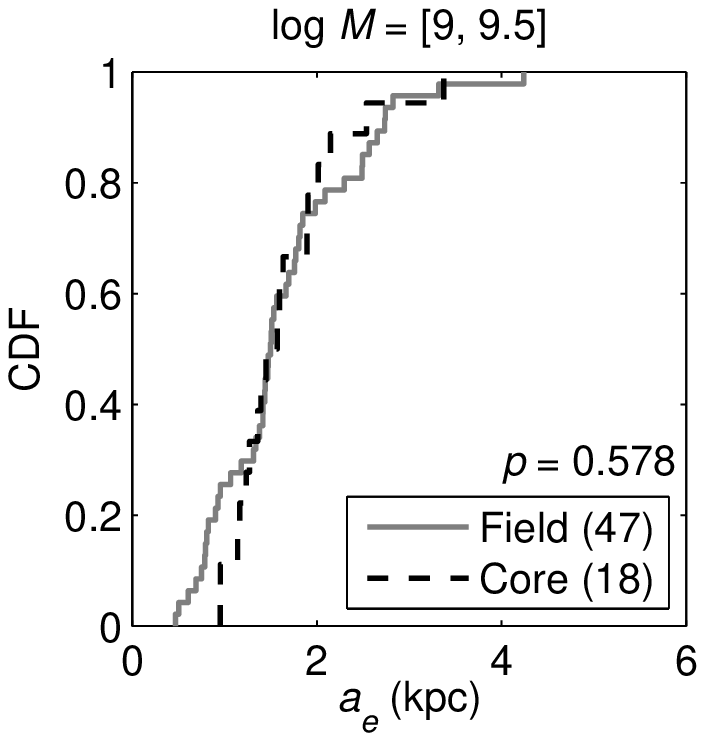}
\includegraphics[width=0.3\textwidth]{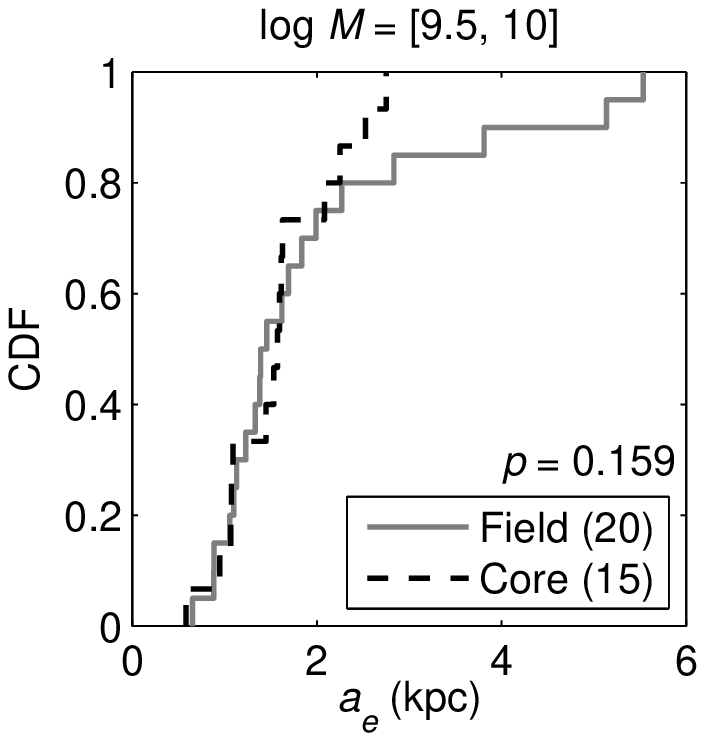} 
\includegraphics[width=0.3\textwidth]{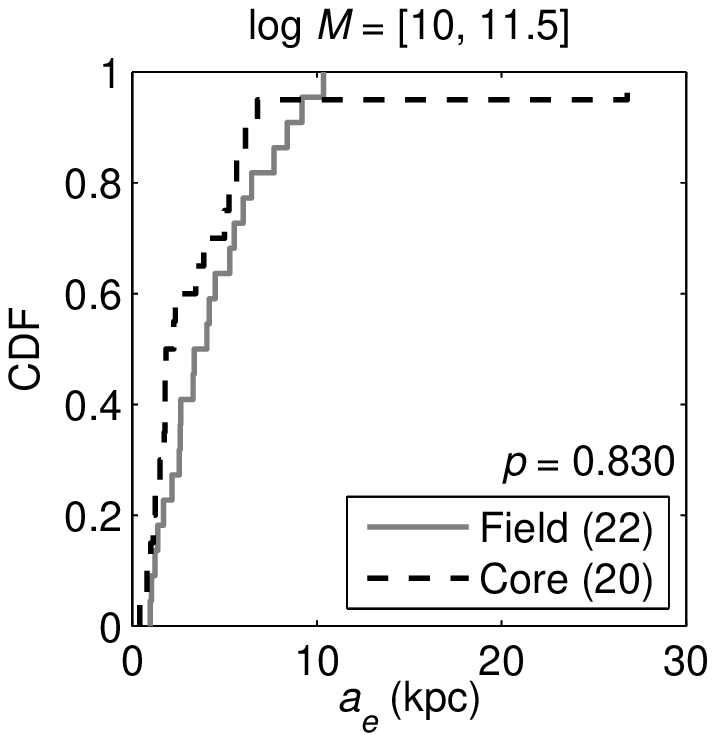} \\
\caption{\label{K-S_E} The $a_e$ cumulative distribution functions (CDFs) for our elliptical galaxies. {\em Top Row:} A comparison of 
the CDFs for elliptical galaxies in the field ({\em grey line}) and the cluster ({\em black dashed line}) for different stellar mass 
bins. {\em Bottom Row:} A similar comparison for elliptical galaxies in the field ({\em grey line}) and cluster core ({\em black 
dashed line}). Respective sample sizes are shown in the legend and the probability $p$ that compared samples are drawn from different 
continuous $a_e$ distributions is shown in the bottom right of each plot. {\em M} is the stellar mass in solar mass units. We find no
significant difference between the CDFs in each environment and no evidence to suggest our elliptical samples are drawn from different
continuous $a_e$ distributions.}
\end{figure*}

\begin{figure*}
\includegraphics[width=0.3\textwidth]{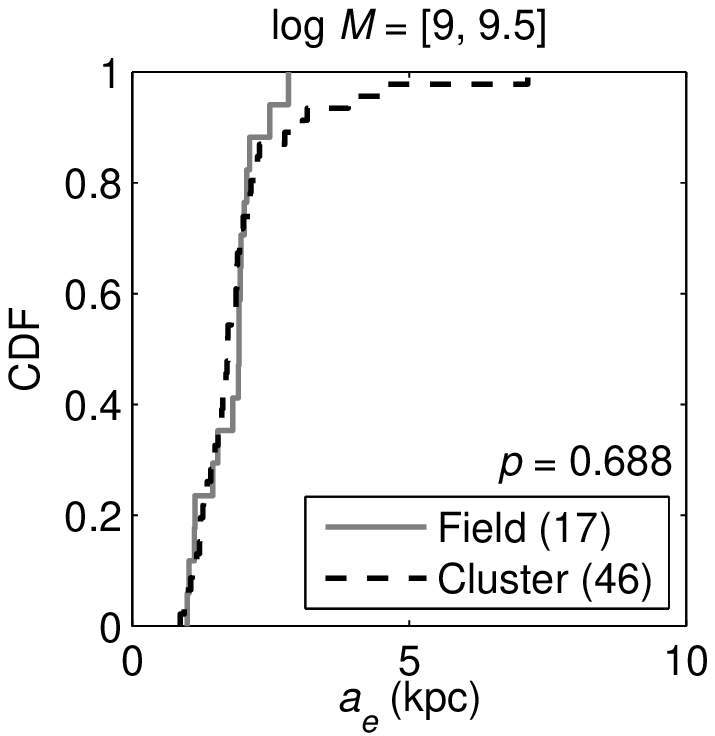}
\includegraphics[width=0.3\textwidth]{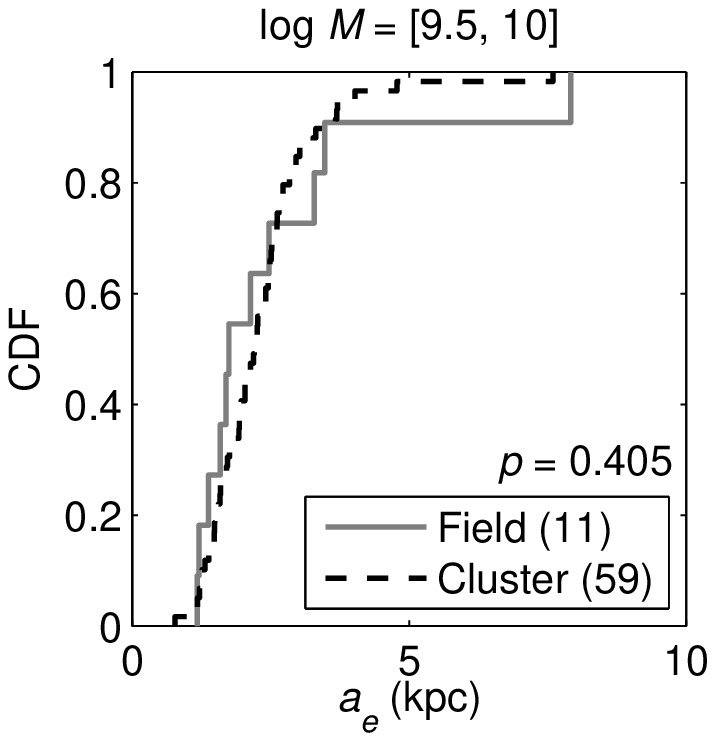}      
\includegraphics[width=0.3\textwidth]{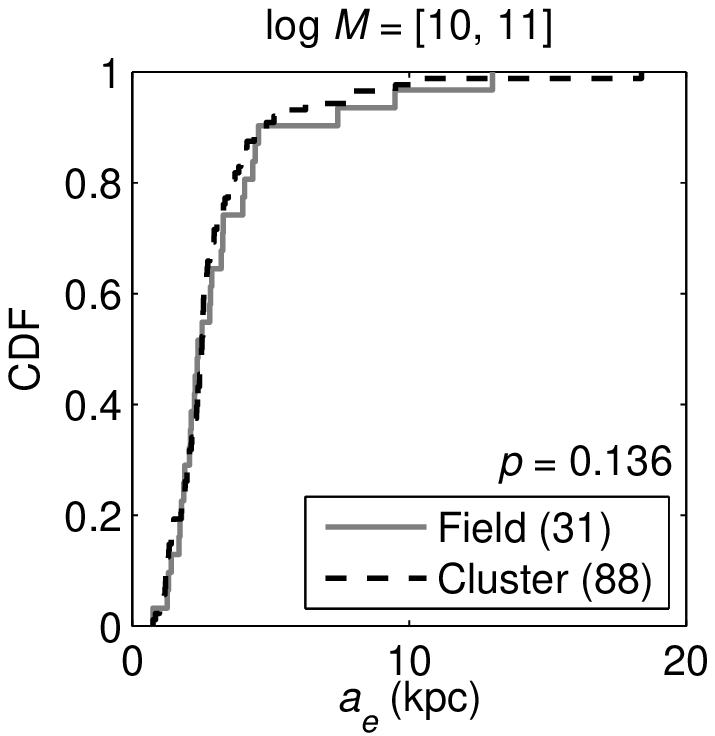}      \\
\includegraphics[width=0.3\textwidth]{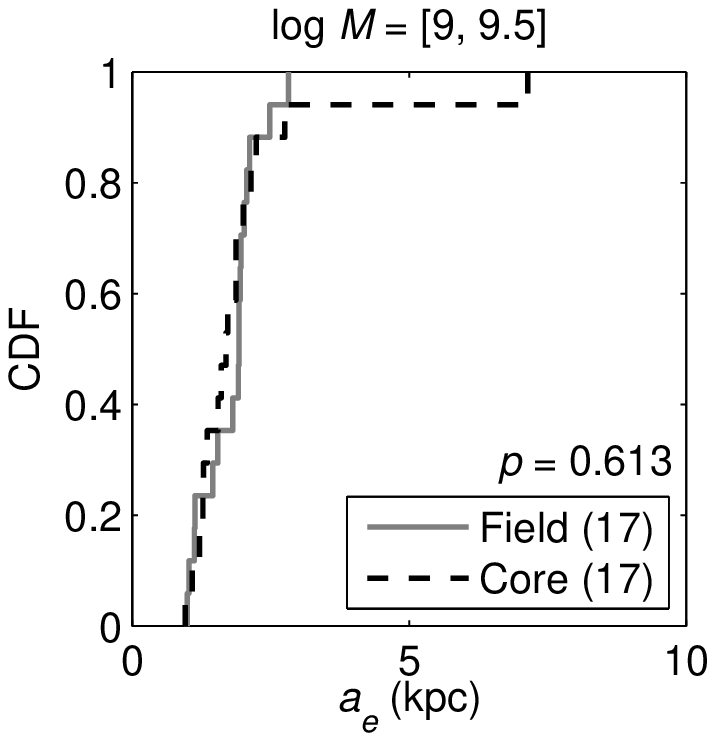}
\includegraphics[width=0.3\textwidth]{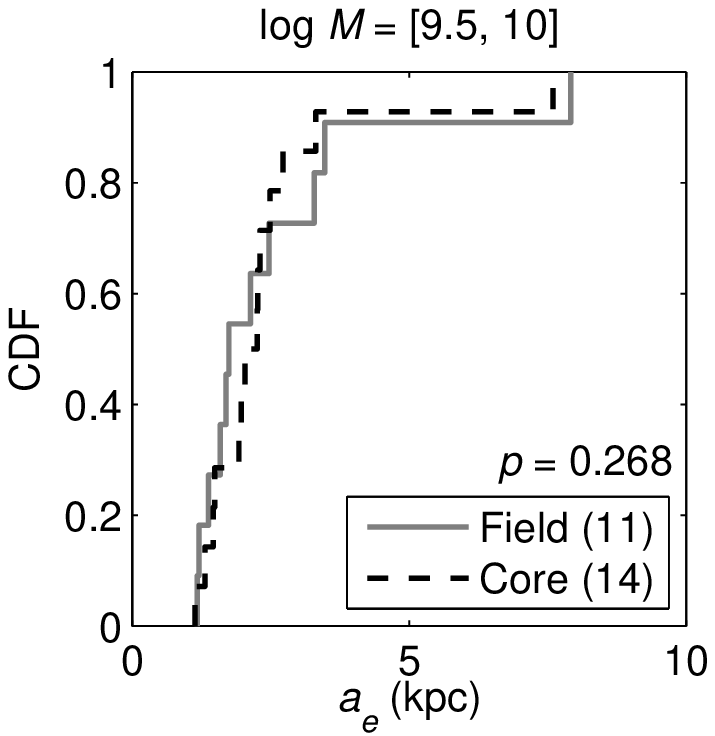} 
\includegraphics[width=0.3\textwidth]{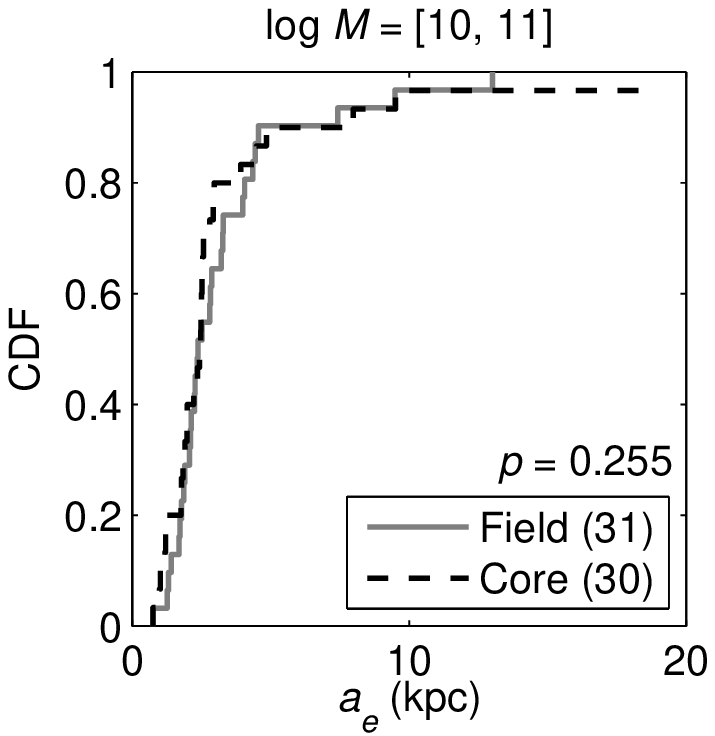} \\
\caption{\label{K-S_S0} The $a_e$ cumulative distribution functions (CDFs) for our lenticular galaxies. {\em Top Row:} A comparison of 
the CDFs for lenticular galaxies in the field ({\em grey line}) and the cluster ({\em black dashed line}) for different stellar mass 
bins. {\em Bottom Row:} A similar comparison for lenticular galaxies in the field ({\em grey line}) and cluster core ({\em black 
dashed line}). Respective sample sizes are shown in the legend and the probability $p$ that compared samples are drawn from different 
continuous $a_e$ distributions is shown in the bottom right of each plot. {\em M} is the stellar mass in solar mass units. We find no
significant difference between the CDFs in each environment and no evidence to suggest our lenticular samples are drawn from different
continuous $a_e$ distributions.}
\end{figure*}

\begin{figure*}
\includegraphics[width=0.3\textwidth]{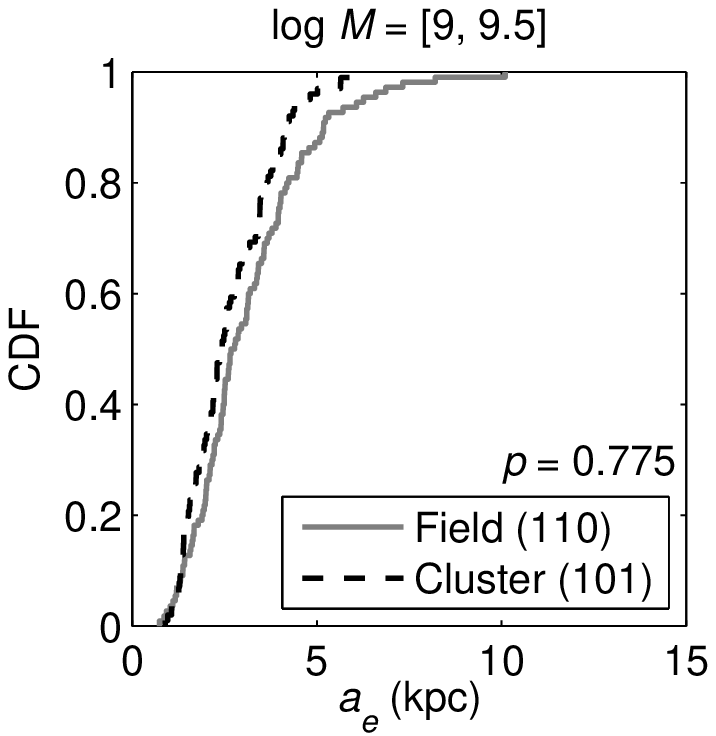}
\includegraphics[width=0.3\textwidth]{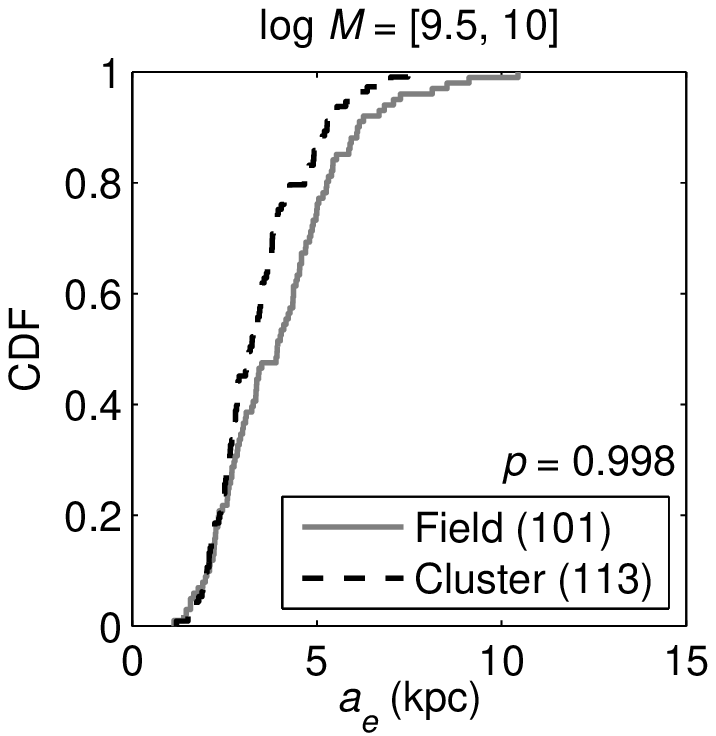}         
\includegraphics[width=0.3\textwidth]{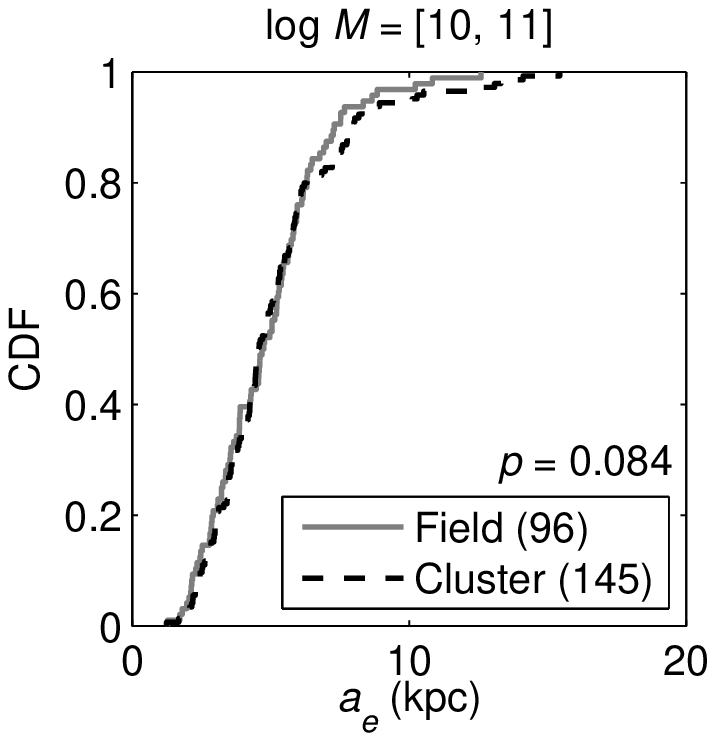}      \\
\includegraphics[width=0.3\textwidth]{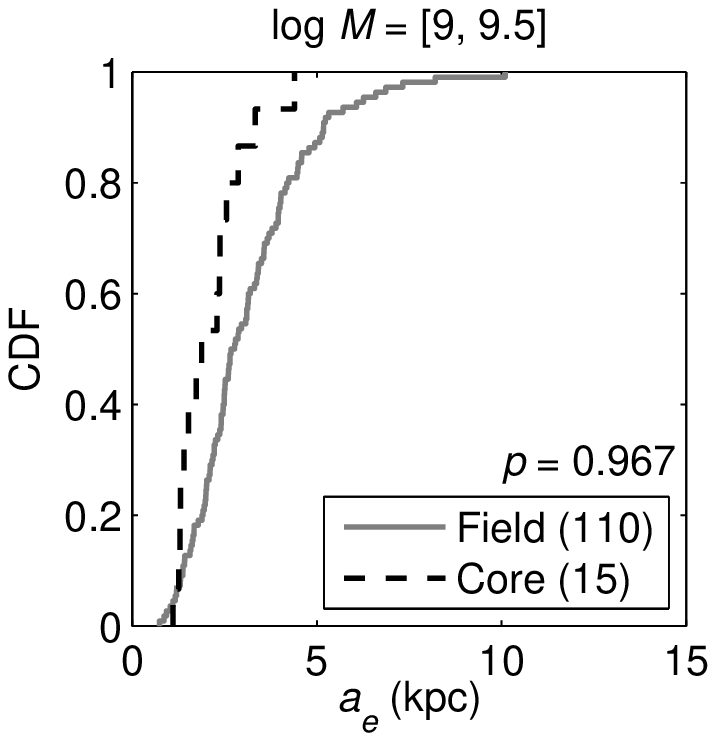}
\includegraphics[width=0.3\textwidth]{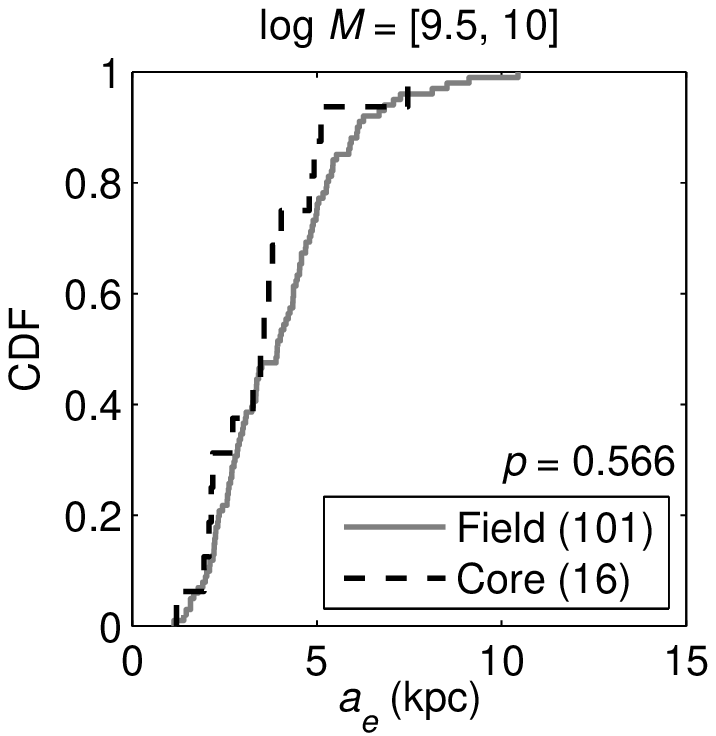} 
\includegraphics[width=0.3\textwidth]{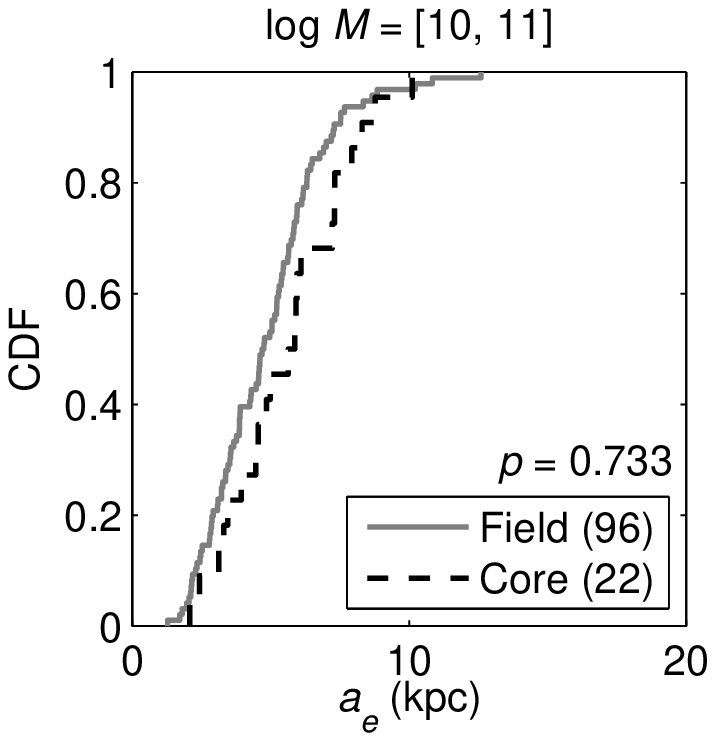} \\
\caption{\label{K-S_Sp} The $a_e$ cumulative distribution functions (CDFs) for our spiral galaxies. {\em Top Row:} A comparison of 
the CDFs for spiral galaxies in the field ({\em grey line}) and the cluster ({\em black dashed line}) for different stellar mass bins. 
{\em Bottom Row:} A similar comparison for spiral galaxies in the field ({\em grey line}) and cluster core ({\em black dashed line}).
Respective sample sizes are shown in the legend and the probability $p$ that compared samples are drawn from different continuous 
$a_e$ distributions is shown in the bottom right of each plot. {\em M} is the stellar mass in solar mass units. For high mass spirals 
(${\rm log}M=[10, 11]$) we find no significant difference between the CDFs in each environment and no evidence to suggest our spiral
samples are being drawn from different continuous $a_e$ distributions. For intermediate-mass spirals (${\rm log}M=[9.5, 10]$) we find 
a significant difference between the CDFs of the field and cluster samples and a high probability (at the $3\sigma$ level) that the 
samples are being drawn from different continuous $a_e$ distributions. However, we find no difference in the comparison of the field 
and the core. For low-mass spirals (${\rm log}M=[9, 9.5]$) we find a significant difference between the CDFs of the field and core 
samples with a high probability (at the $2\sigma$ level) that the samples are being drawn from different continuous $a_e$ 
distributions.} 
\end{figure*}

To test the significance of these results, for each Hubble type and environment we construct $a_e$ cumulative distribution functions 
(CDFs), see Figs.~\ref{K-S_E}--\ref{K-S_Sp}, and perform Kolmogorov--Smirnov (K--S) tests between corresponding mass-selected 
sub-samples from the field and the cluster in order to obtain the probability that the these samples are drawn from different 
continuous $a_e$ distributions. We apply the same K--S tests in a comparison of corresponding mass-selected sub-samples from the field 
and cluster core. The results per mass bin for each Hubble type are shown in Table~\ref{results}.


\begin{table*}
\begin{minipage}{175mm}
\centering
\caption{\label{results} The environmental dependence of the stellar mass-size relation: The mean semi-major axis effective radii 
$\overline{a}_e$ ($\pm1\sigma$) of our elliptical, lenticular, and spiral galaxies in different stellar mass bins for the field, 
cluster, and cluster core environments (in kpc), and the K--S test results for elliptical, lenticular, and spiral galaxies: The 
probability that the field and cluster $p({\rm Field/Cluster})$, and field and cluster core $p({\rm Field/Core})$ samples are being 
drawn from different continuous $a_e$ distributions. We find no environmental effect on $\overline{a}_e$ or environmental dependence 
of the stellar mass-size relation for all morphological types, except for intermediate/low-mass spirals.}
\begin{tabular}{lccccc}
\hline
\hline
{Elliptical Galaxies (E)}            & {}		       & {} 	                   & {}                    & {}	                    & {}         \\
\hline
{${\rm log}M_*/{\rm M_{\odot}}$ range}&{$\overline{a}_e$(Field)}&{$\overline{a}_e$(Cluster)}&{$\overline{a}_e$(Core)}&{$p(\rm Field/Cluster)$}&{$p(\rm Field/Core)$}\\
\hline
{[9, 9.5]}	                     & 1.62 ($\pm$0.11)        & 1.63 ($\pm$0.11)          & 1.64 ($\pm$0.14)      & 0.745	            & 0.578      \\
{[9.5, 10]}		             & 1.96 ($\pm$0.31)        & 1.75 ($\pm$0.12)          & 1.58 ($\pm$0.16)      & 0.554	            & 0.159      \\
{[10, 11.5]}		             & 4.29 ($\pm$0.59)        & 3.62 ($\pm$0.51)          & 4.03 ($\pm$1.28)      & 0.825	            & 0.830      \\
\hline
\hline
{Lenticular Galaxies (S0)}           & {}                      & {}                        & {}                    & {}	                    & {}         \\
\hline
{${\rm log}M_*/{\rm M_{\odot}}$ range}&{$\overline{a}_e$(Field)}&{$\overline{a}_e$(Cluster)}&{$\overline{a}_e$(Core)}&{$p(\rm Field/Cluster)$}&{$p(\rm Field/Core)$} \\
\hline
{[9, 9.5]}		             & 1.78 ($\pm$0.12)        & 1.94 ($\pm$0.16)          & 1.99 ($\pm$0.34)      & 0.688	            & 0.613      \\
{[9.5, 10]}	                     & 2.55 ($\pm$0.59)        & 2.33 ($\pm$0.14)          & 2.45 ($\pm$0.43)      & 0.405	            & 0.268	 \\
{[10, 11]}	                     & 3.27 ($\pm$0.46)        & 3.01 ($\pm$0.25)          & 3.25 ($\pm$0.63)      & 0.136	            & 0.255      \\
\hline
\hline
{Spiral Galaxies (Sp)}               & {} 		       & {}                        & {}                     & {}	            & {}         \\
\hline
{${\rm log}M_*/{\rm M_{\odot}}$ range}&{$\overline{a}_e$(Field)}&{$\overline{a}_e$(Cluster)}&{$\overline{a}_e$(Core)}&{$p(\rm Field/Cluster)$}&{$p(\rm Field/Core)$} \\
\hline
{[9, 9.5]}		             & 3.14 ($\pm$0.16)        & 2.63 ($\pm$0.12)          & 2.11 ($\pm$0.24)      & 0.775	            & 0.967      \\
{[9.5, 10]}		             & 4.00 ($\pm$0.18)        & 3.42 ($\pm$0.12)          & 3.52 ($\pm$0.39)      & 0.998	            & 0.566      \\
{[10, 11]}		             & 4.85 ($\pm$0.21)        & 5.10 ($\pm$0.21)          & 5.61 ($\pm$0.46)      & 0.084	            & 0.733      \\
\hline
\hline
\end{tabular}
\end{minipage}
\end{table*}

In this work, we only consider an environmental effect on the stellar mass-size relation to be significant if K--S tests yield a
$2\sigma$ level probability for the field and cluster/core samples being drawn from different continuous $a_e$ distributions.

For elliptical and lenticular galaxies, the probability that the field and cluster $p({\rm Field/Cluster})$, and the field and cluster
core $p({\rm Field/Core})$ samples are drawn from different continuous $a_e$ distributions is below the $2\sigma$ level in all mass 
bins. Therefore, we find no evidence to suggest an environmental dependence on the stellar mass-size relation for these galaxies. This 
result is in qualitative agreement with the work of \cite{Rettura_etal:2008} who came to the same conclusion at higher redshift 
($z\sim1$) using a much smaller sample of 45 cluster and field early-type galaxies.

For high-mass spiral galaxies (${\rm log} M_*/{\rm M_{\odot}}=[10,11]$), both the probabilities $p({\rm Field/Cluster})$ and $p({\rm
Field/Core})$ are below the $2\sigma$ level. Therefore, we find no evidence to suggest an environmental dependence on the stellar 
mass-size relation for high-mass spirals. However, for intermediate and low-mass spirals we find evidence for a potential environmental 
dependence on the stellar mass-size relation. For intermediate-mass spirals (${\rm log}M_*/{\rm M_{\odot}}=[9.5,10]$), the probability 
$p({\rm Field/Core})$ is below the $1\sigma$ level, however, we find a significant probability (at the $3\sigma$ level) for $p({\rm 
Field/Cluster})$. This is in agreement with the difference seen between $\overline{a}_e$ for intermediate-mass spirals in the field and 
cluster environments, see Fig.~\ref{M-S}, where we find a larger $\overline{a}_e$ for field spirals. For low-mass spirals (${\rm log} 
M_*/{\rm M_{\odot}}=[9,9.5]$), the probability $p({\rm Field/Cluster})$ is below the $2\sigma$ level, however, $p({\rm Field/Core})$ is 
at the $2\sigma$ level. Again this is in agreement with the difference seen between $\overline{a}_e$ for low-mass spirals in the field 
and core environments, see Fig.~\ref{M-S}, where we find a larger $\overline{a}_e$ for field spirals. Therefore, there appears to be a 
population of large-$a_e$, intermediate/low-mass spirals in the field that is largely absent from the cluster and cluster core 
environments.


\section[]{Discussion}

\label{Discussion}

The main result of this work is the lack of any environmental dependence for the majority of the galaxy types studied. The 
significance observed for intermediate/low-mass spirals is inconclusive and requires further verification by similar studies in other
fields. In this section we discuss some of the considerations for the reliability of our results and for establishing the stellar 
mass-size relation at low stellar masses. 

The majority of our field galaxies have redshift $z\sim0.25$ while the cluster galaxies have $z\sim0.17$. Using the fits of
\cite{Buitrago_etal:2008}, we evaluated the expected size evolution between the mean redshifts of our field and cluster samples 
to be $\sim10$ per cent ($5.6$ per cent for discs and $10.3$ per cent for spheroids). Presently, size evolution has only been 
characterised for the most massive galaxies \citep[$M_*>10^{11}{\rm M_{\odot}}$;][]{Trujillo_etal:2007, Buitrago_etal:2008}. 
Consequently, these fits do not hold for our galaxy samples (mostly with $M_*<10^{11}{\rm M_{\odot}}$) but we use them to give an idea 
of the degree of expected evolution. Therefore, neglecting any environmental effects, on average our field galaxies are expected to 
be $\sim10$ per cent smaller than the cluster galaxies. However, we observe the opposite trend for our spiral galaxies at 
intermediate/low stellar masses.

The evidence we find for an environmental dependence on the stellar mass-size relation for intermediate/low-mass spirals does 
not however appear in a consistent way. The significance of $p(\rm Field/Cluster)$ is at the $3\sigma$ level for intermediate masses 
(${\rm log} M_*/{\rm M_{\odot}=[9.5,10]}$), but only just above the $1\sigma$ level for low masses (${\rm log} M_*/{\rm M_{\odot}=
[9,9.5]}$). Either the significance at intermediate masses is not robust and there is no environmental dependence or the significance 
at low masses is being masked by some effects (e.g. photometric redshift errors and contaminations). We explore these effects in the 
following sections.


\subsection[]{The Effect of Photometric Redshift Errors}

\label{photo-z_errors}

\begin{figure*}
\includegraphics[width=0.49\textwidth]{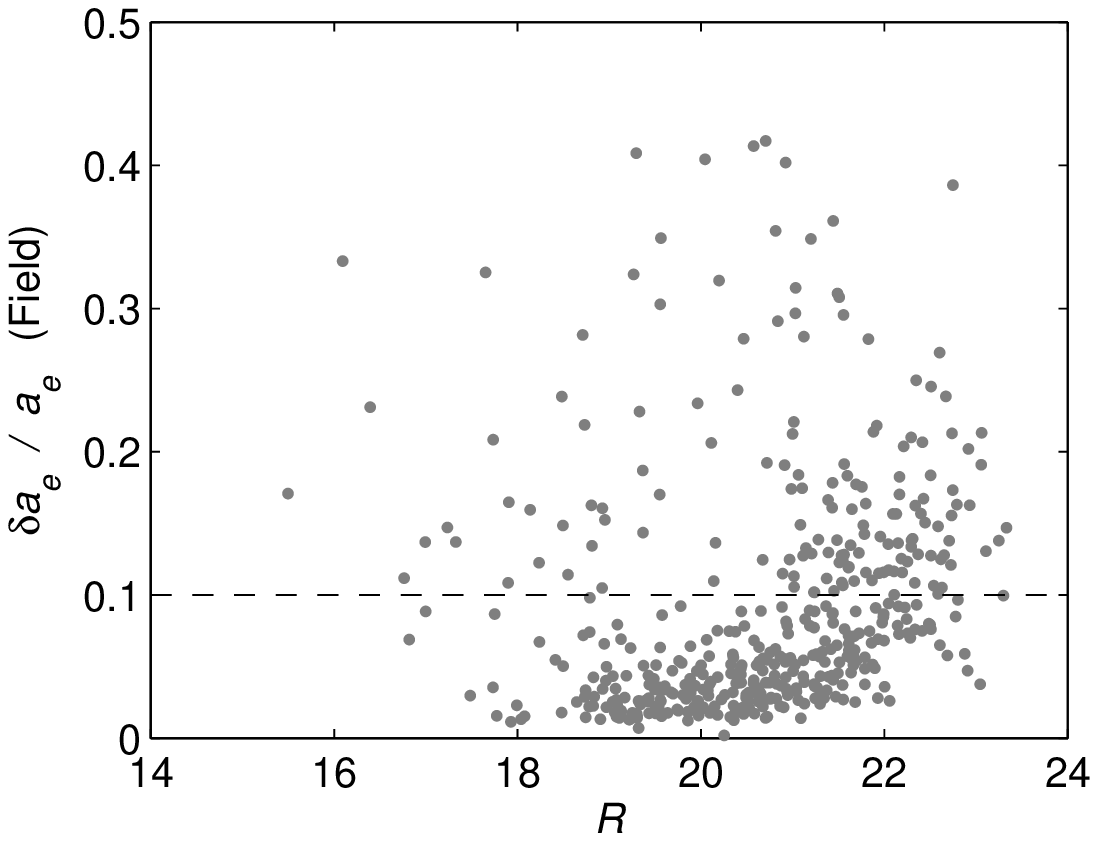}
\includegraphics[width=0.49\textwidth]{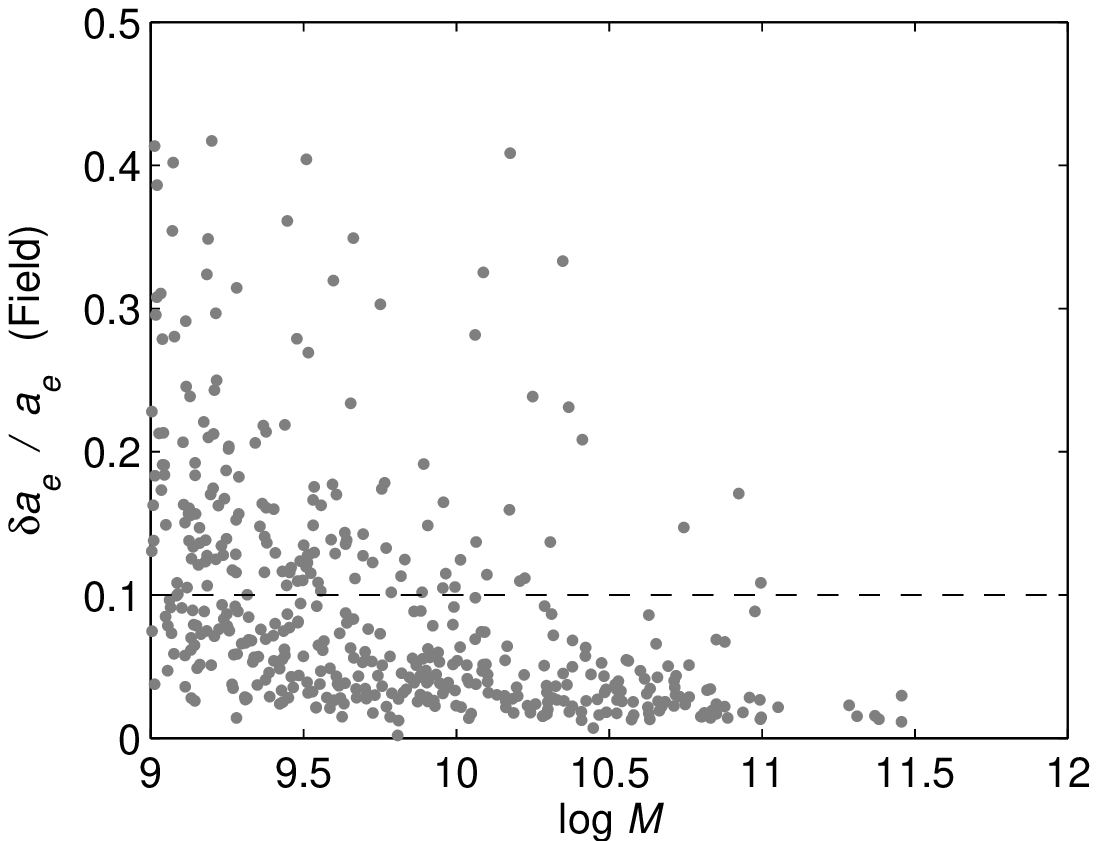}
\caption{\label{D_err} {\em Left panel:} The relative error in the physical size $a_e$ associated with the photo-$z$ error 
$\delta{a_e}/a_e$ {\bf (i.e. error in distance to galaxy)} for our field sample vs. $R$-band magnitude showing that the error increases 
towards fainter magnitudes due to the increasing photo-$z$ error. {\em Right panel:} $\delta{a_e}/a_e$ for our field sample vs. stellar 
mass showing that the error increases with decreasing stellar mass. To test the robustness of our results we discard all high error 
objects ($>10$ per cent) ({\em above dashed line}) from our field sample and assess the effect on our K--S test results.} 
\end{figure*}

In the determination of our physical galaxy sizes $a_e$ we have used the cluster redshift ($z = 0.167$) for our cluster samples but 
used the original COMBO-17 redshift estimates for our field sample. Therefore, the errors in the COMBO-17 photometric redshifts 
{\bf($\delta{z}$)} propagate into the physical size measurements of our field galaxies. Fig.~\ref{D_err} shows the relative error in 
$a_e$ associated with this photo-$z$ error (i.e. distance to galaxy ($D$) error) $\delta{a_e}/a_e = {\rm max} 
|D(z) - D(z\pm\delta{z})| /D(z)$, plotted against $R$-magnitude and stellar mass for our field galaxies. The errors can reach out to 
$\sim50$ per cent on a galaxy-galaxy basis and generally increase towards fainter $R$-magnitudes and with decreasing stellar mass. We 
therefore need to be careful that large $\delta{a_e}/a_e$ galaxies are not affecting our analysis, especially at intermediate/low 
stellar masses where we find a difference in $\overline{a}_e$ between the field and cluster/core spirals.

\begin{figure}
\includegraphics[width=0.49\textwidth]{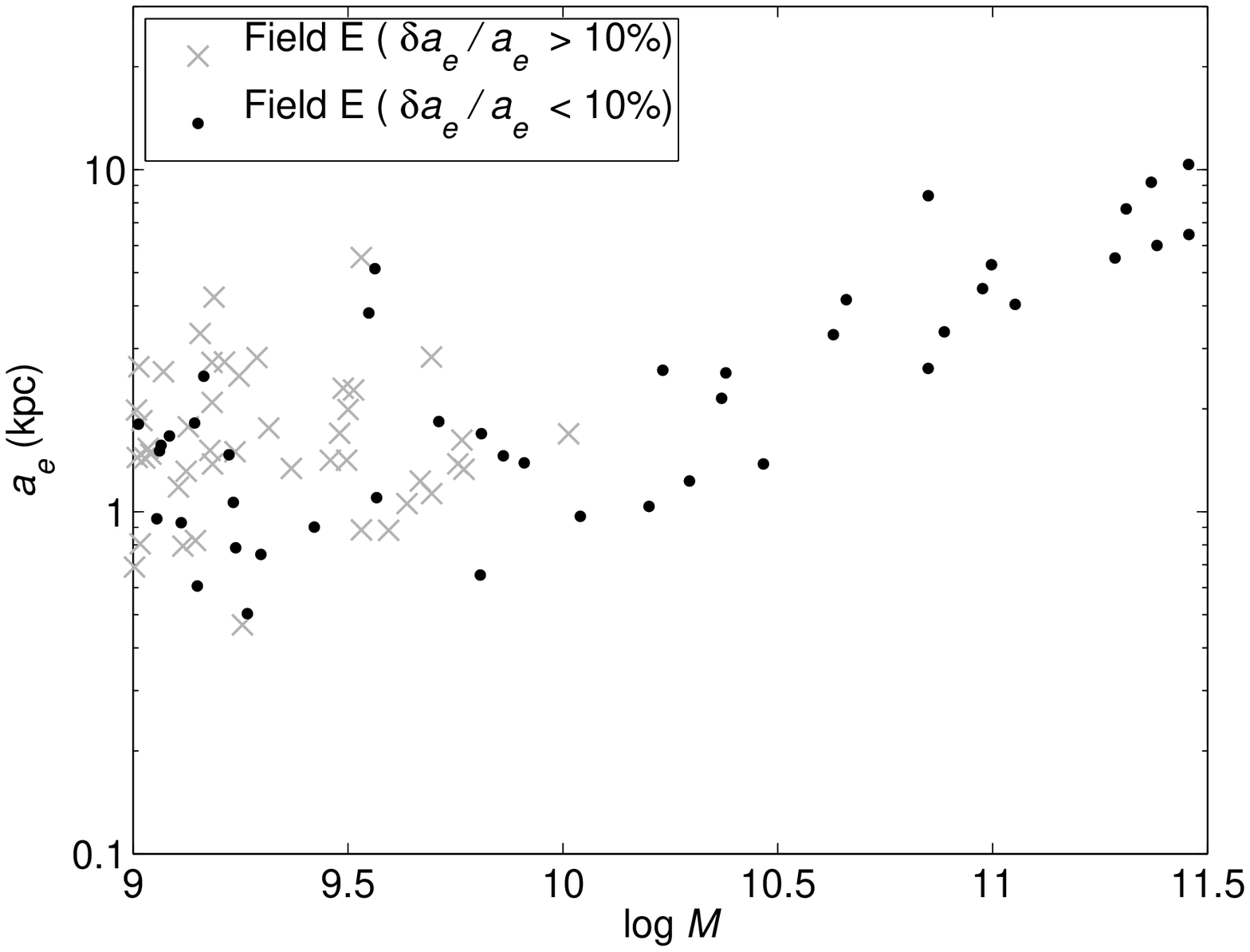}  
\includegraphics[width=0.49\textwidth]{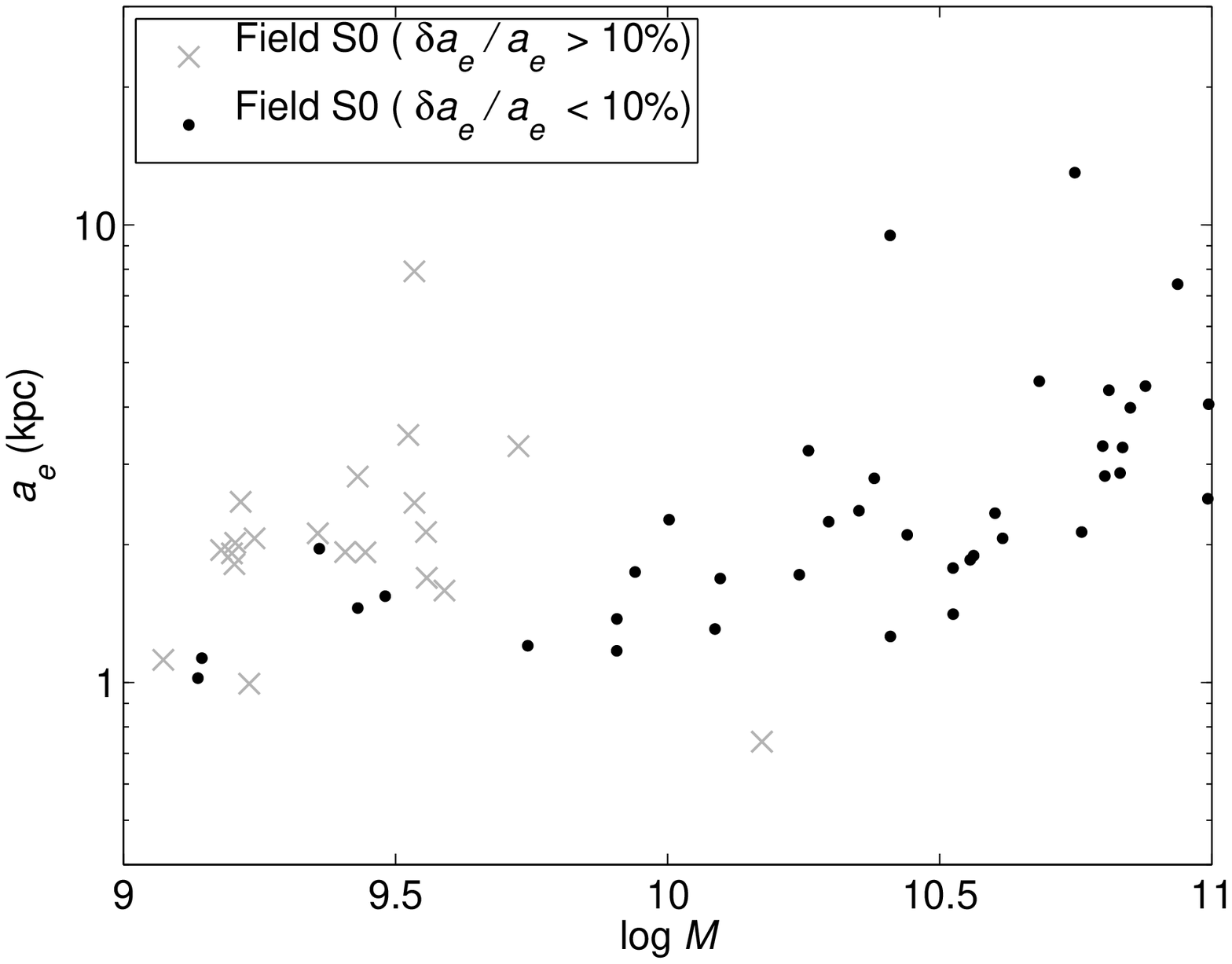}
\includegraphics[width=0.49\textwidth]{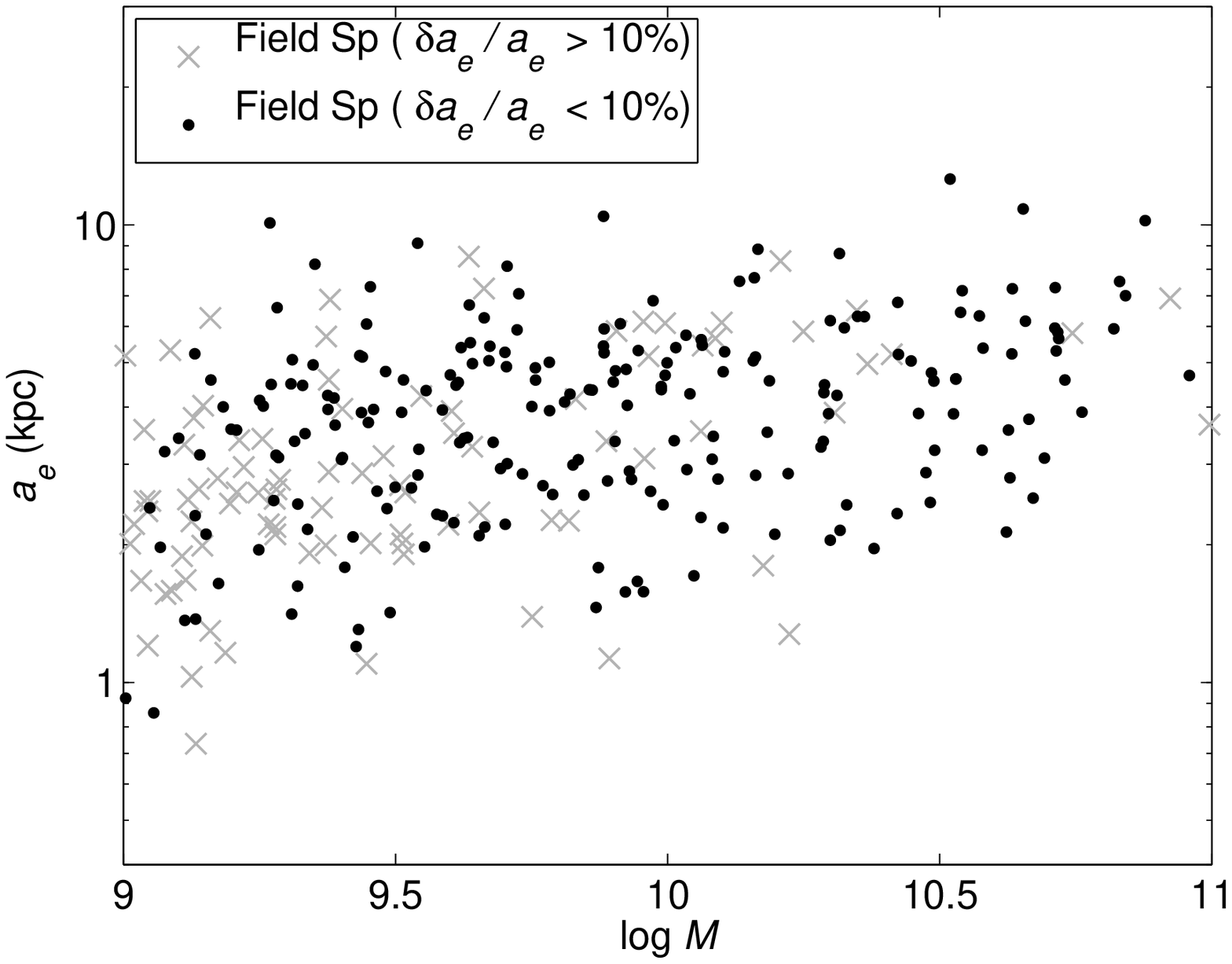} 
\caption{\label{selection_effect} Testing the effects of large $\delta{a_e}/a_e$ (distance) errors. The stellar mass-size 
relations for our field samples showing galaxies with a relative error $\delta{a_e}/a_e>10$ per cent ({\em grey}) and 
$\delta{a_e}/a_e<10$ per cent ({\em black}) for elliptical ({\em top}), lenticular ({\em centre}), and spiral ({\em bottom}) galaxies, 
respectively.}
\end{figure}

The individual COMBO-17 photo-$z$ errors are expected errors based on the shape of the redshift probability distribution ($z$PDF) of 
a galaxy. As such they can sometimes underestimate the true error when the $z$PDF is based on an incomplete set of comparison 
templates (i.e. the real spectrum is an unusual one). When the $z$PDF is wide and the estimated error is large, it means that
redshift ambiguities are actually present. It is a true sign that the redshift is potentially unreliable. 
In order to assess the effect of large $\delta{a_e}/a_e$ galaxies on our mass-size relations and analysis, we remove all galaxies 
from our field sample with $\delta{a_e}/a_e>10$ per cent, see Figs.~\ref{D_err}--\ref{selection_effect}.

For elliptical galaxies there was no effect on the level of significance of the K--S test results; $p({\rm Field/Cluster})$ and
$p({\rm Field/Core})$, in each mass bin. In no cases did the level of significance go above the $2\sigma$ level. Therefore, our
conclusion that there is no evidence to suggest an environmental dependence on the stellar mass-size relation for elliptical galaxies
is robust with respect to large $\delta{a_e}/a_e$ galaxies.

For lenticular galaxies there was no effect on the level of significance of the K--S test results for high and low masses
(${\rm log}M_*/{\rm M_{\odot}}=[9, 9.5][10, 11]$) with $p({\rm Field/Cluster})$ and $p({\rm Field/Core})$ remaining below the 
$2\sigma$ significance level. However, at intermediate masses (${\rm log}M_*/{\rm M_{\odot}}=[9.5, 10]$) both $p({\rm Field/Cluster})$ 
and $p({\rm Field/Core})$ increase in significance to the $2\sigma$ level. In this case, the removal of the large  $\delta{a_e}/a_e$
galaxies has introduced a selection effect (see Fig.~\ref{selection_effect}) causing a bias towards smaller sizes and also 
dramatically reduced the number of field lenticulars in this mass bin. Therefore, we do not trust the high significance in this case 
and our conclusion that there is no evidence to suggest an environmental dependence on the stellar mass-size relation for these 
galaxies is robust with respect to large $\delta{a_e}/a_e$ galaxies.

For high-mass spirals (${\rm log}M_*/{\rm M_{\odot}}=[10,11]$) the significance is also robust with respect to the removal of the large
$\delta{a_e}/a_e$ galaxies and is below the $2\sigma$ level for both the $p({\rm Field/Cluster})$, and $p({\rm Field/Core})$. For 
intermediate-mass spirals (${\rm log}M_*/{\rm M_{\odot}}=[9.5,10]$) the high significance of $p({\rm Field/Cluster})$ increases to above 
the $3\sigma$ level, however, $p({\rm Field/Core})$ remains below the $2\sigma$ level, possibly due to low number statistics. For 
low-mass spirals (${\rm log}M_*/{\rm M_{\odot}}=[9,9.5]$) the high significance of $p({\rm Field/Core})$, remains at the $2\sigma$ level, 
and the significance of $p({\rm Field/Cluster})$ increases from below the $2\sigma$ level to the $3\sigma$ level, indicating that large 
$\delta{a_e}/a_e$ errors could be masking the effect in this region. Therefore, our conclusion that the field and cluster/core samples 
are potentially being drawn from different continuous $a_e$ distributions at intermediate/low-masses is robust with respect to large 
$\delta{a_e}/a_e$ errors.

\subsection[]{Photometric Redshift Bias}

\label{Photometric_Redshift_Bias}

At present, photometric redshifts on the STAGES (A901/2) field have only been assessed for quality by comparison to spectroscopic 
redshifts at bright magnitudes. \cite{Wolf_etal:2004} find for a sample of 404 bright galaxies with $R<20$ and $z = [0,0.3]$, 351 of 
which were from the A901/2 field, that $77$ per cent had photo-$z$ deviations of $|\delta_z/(1+z)|<0.01$, and only $1$ per cent had 
deviations $|\delta_z/(1+z)|>0.04$ from the true redshift.

There are currently no spectroscopic redshifts on the A901/2 field for comparison at faint magnitudes. However, a faint spectroscopic 
dataset does exist on the COMBO-17 CDFS field from the VIMOS VLT Deep Survey \citep[VVDS;][]{LeFevre_etal:2004}. 
\cite{Hildebrandt_etal:2008} use a sample of 420 high-quality spectroscopic redshifts that are reasonably complete to $R_{\rm{ap}}<23$ 
and find a $1\sigma$ scatter of 0.018 in $\delta{z}/(1+z)$ and also a mean bias of $-0.011$.

The mean redshift bias is dependent on errors in the photometric calibration. In COMBO-17 the photometric calibration has slight
variations from field to field and therefore the photo-$z$ bias measured on the CDFS cannot be assumed to hold for the A901/2 field. 
However, we need to be careful that any environmental effect on the stellar mass-size relation between the field and cluster/core 
samples is not due to a redshift bias on our field sample causing biased galaxy $a_e$. In order to address this, we use the bias for 
the CDFS field as an order of magnitude estimate for any potential redshift bias on the A901/2 field. We then assess the degree of 
uncertainty caused by a $\pm$ redshift bias on our field sample, and evaluate its propagation into our results. We do not consider 
the result valid if the redshift bias introduces any significance at the $2\sigma$ level due to the uncertain nature of the bias. 
However, we test to see if the bias causes our high significance results (i.e. intermediate/low-mass spirals) to drop below the 
$2\sigma$ level.

For intermediate mass spirals (${\rm log}M_*/{\rm M_{\odot}}=[9.5,10]$), $p({\rm Field/Cluster})$ is only robust to a positive redshift 
bias. However, after the removal of large $\delta{a_e}/a_e$ galaxies the level of significance is completely robust to a $\pm$ redshift 
bias and is always above the $3\sigma$ level. For low-mass spirals (${\rm log}M_*/{\rm M_{\odot}}=[9,9.5]$), the level of significance 
for $p({\rm Field/Core})$ is only robust to a positive redshift bias. However, after the removal of large $\delta{a_e}/a_e$ galaxies 
the level of significance is completely robust to a  $\pm$ redshift bias and is always above the $3\sigma$ level. For 
$p({\rm Field/Cluster})$ after the removal of large $\delta{a_e}/a_e$ galaxies the significance is completely robust to a $\pm$ 
redshift bias and is always above the $2\sigma$ level.


\subsection[]{The Effect of Contaminations}

Contamination of cluster samples by the field could also have potentially masked an environmental dependence on the stellar 
mass-size relation in our comparison of the field and cluster for low-mass spirals. In order to address this, we use Monte-Carlo 
simulations to evaluate the effect of contaminations on $\overline{a}_e$ in each stellar mass bin for spiral galaxies.

In each environment and stellar mass bin we assume the true $a_e$ distribution of spiral galaxies follows a Gaussian with the same 
mean and standard deviation as the corresponding sub-sample in the actual data. We then use Monte-Carlo simulations to sample 
respective $a_e$ distributions in order to obtain a simulated sample of field and ``pure'' cluster galaxies in each stellar mass bin. 
We then perform K--S tests between corresponding sub-samples from the field and cluster distributions in order to obtain the 
probability $p({\rm Field/Cluster})$ in each stellar mass bin. The pure cluster samples were then contaminated by galaxies sampled 
from the field distribution, with a fraction corresponding to the level given in Table~\ref{tbl1}. Respective sample sizes were matched 
to that of the actual data. The same K--S tests were then applied between the field and contaminated cluster samples. This Monte-Carlo 
simulation process was repeated 100,000 times.

For each iteration we also obtain $\Delta\overline{a}_e$ between the pure and contaminated cluster samples in each stellar mass 
bin in order to assess if any potential bias in $\overline{a}_e$ is caused by the contamination by the field. We then correct for this 
bias and evaluate its effect on our K--S tests in order to assess if the contamination is potentially masking an environmental 
dependence for spiral galaxies at low stellar masses. For high and intermediate masses (${\rm log}M_*/{\rm M_{\odot}} = [9.5,10],
[10,11]$), this bias has no effect on the significance of the K--S tests. However, we do observe a change in significance at low 
stellar masses (${\rm log}M_*/{\rm M_{\odot}} = [9,9.5]$), where we find that a mean $\overline{a}_e$ bias of $+0.128\,{\rm kpc}$ is 
expected on the cluster sample. Correcting for this bias causes the significance of $p({\rm Field/Cluster})$ to increase from $0.775$ 
to above the $2\sigma$ level ($0.965$). This indicates that the $25$ per cent contamination of the cluster sample by the field is 
potentially masking an environmental dependence at low stellar masses.



\subsection[]{Visual Inspection}

\label{images}

\begin{figure*}
\centering
\includegraphics[width=1\textwidth]{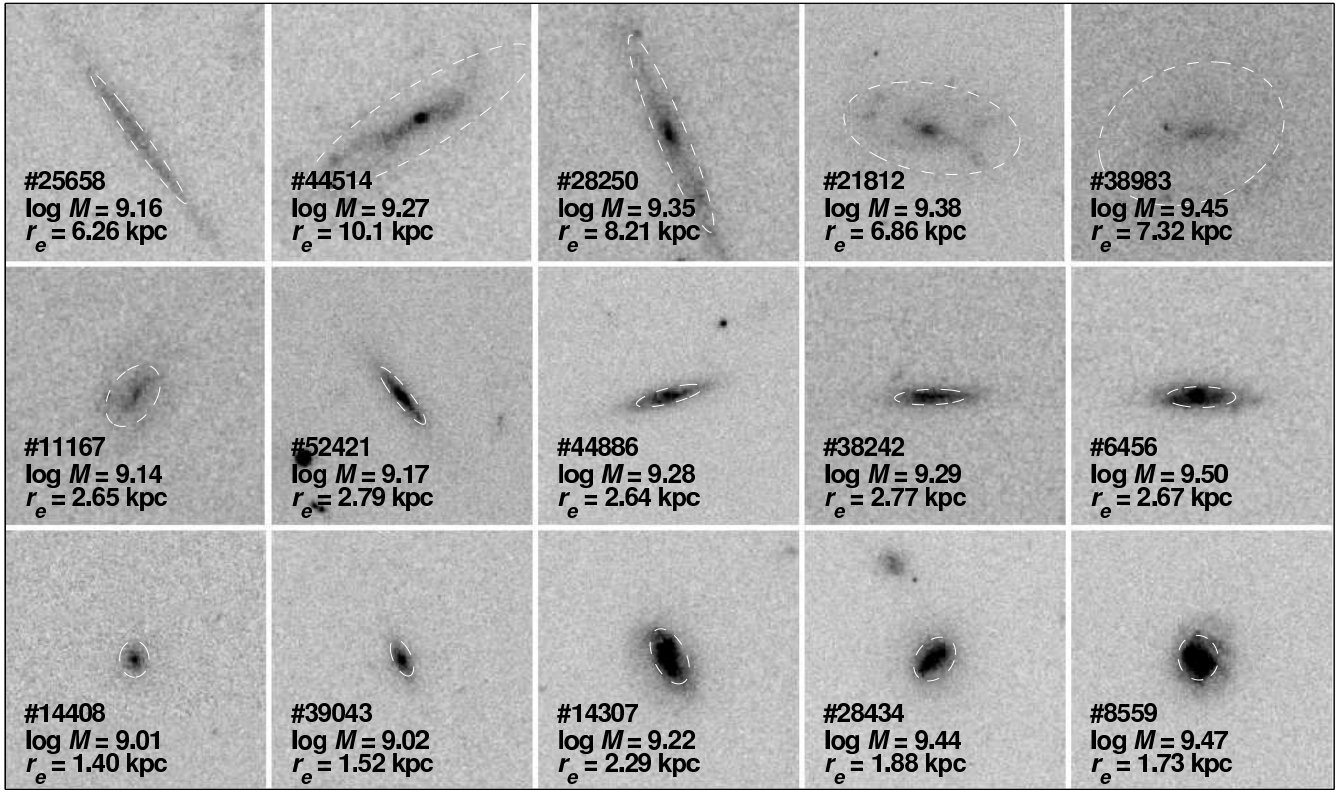}
\caption{\label{Sp_images} Low-mass, ${\rm log}M_*/{\rm M_{\odot}}=[9,9.5]$, spiral galaxies from different environments. {\em Top Row:} 
A selection of large-$a_e$, low-mass spiral galaxies from the field environment. {\em Middle Row:} Median-$a_e$ field spirals of similar 
stellar mass. {\em Bottom Row:} Median-$a_e$ cluster core spiral galaxies of similar stellar mass. The ${a}_e$ isophote is 
over-plotted on each image to give an assessment of physical size of the galaxy. The large-$a_e$ field spirals show the presence of an 
extended disk that is absent from the cluster core spirals. We hypothesize that the extended disk of these large-$a_e$ spirals may not 
survive the environmental conditions of the cluster core. Image size $20\,\rm{kpc}\times20\,\rm{kpc}$.}
\end{figure*}

In order to further validate our results, we examined the ACS images of the low-mass (${\rm log} M_*/{\rm M_{\odot}}=[9,9.5]$), 
large-$a_e$ field spirals and compared them to the ACS images of median-$a_e$ field and cluster core spirals of similar stellar mass. 
The semi-major axis effective radius $a_e$ isophote (calculated using the ellipticity and position angle from the GALFIT model) was
over-plotted on each ACS image in order to assess the physical size of the galaxy. Fig.~\ref{Sp_images} shows a selection of these
comparisons across the low-mass range. We find that the large-$a_e$ field spirals appear to have extended discs that are not present 
in the cluster core spirals. We inspected the ACS images for all 15 low-mass cluster core spirals and find that these extended discs 
are absent from all but one of the low-mass spirals in the cluster core. The exception could well be a field contaminate (we expect a 
$25$ per cent contamination of the cluster sample by the field, see \S~\ref{sample_completeness}) or could be just arriving in the 
core of the cluster. In Fig.~\ref{Sp_images} a couple of the large-$a_e$ spiral fits (e.g. $\#21812$ and $\#38983$) are potentially 
unreliable with the effective radius appearing to be over-estimated. However, the fact that no similar potentially unreliable fits are 
observed in the cluster core suggests the field sample is drawn from a different parent population for low-mass spirals. We hypothesize 
that the extended disks present in the low-mass, large-$a_e$, field spirals do not survive the environmental conditions of the cluster 
core. 

\section[]{Conclusions}

\label{conclusions}

We present the stellar mass-size relations for elliptical, lenticular, and spiral galaxies in three different environments: the field, 
cluster, and cluster core, using {\em HST}/ACS imaging and data from STAGES. We use a large sample of field and cluster galaxies, and 
a cluster sub-sample of cluster core galaxies, and determine the significance of an environmental dependence on the stellar mass-size
relation for each morphological type.

For elliptical and lenticular galaxies, significance of an environmental dependence on the stellar mass-size relation is below the 
$2\sigma$ level for all mass bins, ${\rm log}M_*/{\rm M_{\odot}} = [9,9.5],[9.5,10],[10,11.5]/[10,11]$. These results have been shown 
to be completely robust to galaxies with high photo-$z$ errors and to the effect of a potential photo-$z$ bias on the A901/2 field. 
We conclude that there is no evidence to suggest a significant dependence on the environment of the stellar mass-size relations for 
elliptical and lenticular galaxies. This implies that elliptical and lenticular galaxies have undergone a similar size evolution in 
these environments. This result is in qualitative agreement with the work of \cite{Rettura_etal:2008} who came to the same conclusion 
at higher redshift ($z\sim1$) using a much smaller sample of 45 cluster and field massive ($M_*>5 \times 10^{10}{\rm M_{\odot}}$) 
early-type galaxies.

Size evoution is observed to be stronger for early-type galaxies than late-type galaxies \citep{Trujillo_etal:2007}. Therefore, if the 
environment was the principal driver of size evolution, one would expect the environmental dependence of the stellar mass-size relation 
to be strongest for elliptical and lenticular galaxies. The fact that we find no such dependence suggests internal drivers \citep[e.g., 
AGN feedback,][]{Fan_etal:2008} are governing their size evolution.

For spiral galaxies, we also find no evidence to suggest an environmental dependence on the stellar mass-size relation for high-masses 
(${\rm log} M_*/{\rm M_{\odot}}>10$), with the significance of an environmental dependence below the $2\sigma$ level. However, we find a 
significant environmental dependence (at the $2\sigma$ level) at intermediate/low stellar masses (${\rm log} M_*/{\rm M_{\odot}}<10$). 
We also find the mean $a_e$ for low-mass spirals to be larger in the field than the cluster/core due to a population of large-$a_e$ 
spirals in the field that are largely absent from the cluster/core environments. Inspection of the ACS images of low-mass spirals has 
shown that large-$a_e$ spirals in the field contain extended discs that are largely absent from the cluster and cluster core. We 
hypothesise that the observed fragile discs of these low-mass field spirals may not survive the conditions of the cluster environments. 
The significance of these results have been shown to be completely robust to large photo-$z$ errors and a potential photo-$z$ bias on 
the A901/2 field. This result is in qualitative agreement with the work of \cite{Guo_etal:2009} and \cite{Weinmann_etal:2009}, who find 
that at fixed mass, low-mass late-type satellite galaxies (in larger groups/clusters) are smaller than similar mass late-type central 
galaxies (in smaller groups i.e. the field). 

Size evolution is observed to be stronger for more massive galaxies \citep{Franx_etal:2008}. Therefore, if the environment was the 
principal driver behind the size evolution of spiral galaxies, one would expect the environmental dependence of the stellar mass-size 
relation to be stronger for high-mass spirals. However, the fact that we find no such dependence suggests internal drivers are 
governing their size evolution. For low-mass spirals we do observe a significant effect of the environment on the stellar mass-size 
relation suggesting that the environment probably has an effect on the size evolution at intermediate/low masses.

The results presented here are for one survey field and one multi-cluster complex, therefore, we intend to test the environmental 
dependence of the stellar mass-size relation in other fields and across a wide redshift range. Studies of field and cluster galaxy 
sizes over a wide redshift range will allow galaxy size evolution to be characterised for the field and cluster environments. This 
test will provide the strong evidence needed to assess if the environment really does play a role in galaxy size evolution. Also, in 
order to assess the potential mechanisms responsible for size evolution we intend to extend our studies to look at the environmental 
dependence of galaxy radial surface brightness profiles in STAGES galaxies. By looking at a galaxy radial surface brightness profile, 
we can assess if elliptical and lenticular galaxies have compact cores or cusps, if compact cores are identified it would be evidence 
for smooth envelope accretion (minor/major mergers) having an effect on the size evolution of galaxies. We can also assess if spiral 
galaxies have discs that are truncated or anti-truncated, which will effect the effective radius, and may also be dependent on the 
environment.


\section[]{Acknowledgements}

Support for STAGES was provided by NASA through GO-10395 from STScI operated by AURA under NAS5-26555. DTM acknowledges the support 
of STFC. MEG and CW were supported by an STFC Advanced Fellowship. BH was supported by STFC. MB and EvK acknowledge the support of
FWF grant P18416.


\bibliographystyle{mn2e} \bibliography{DTM_bibtex} \bsp

\label{lastpage}

\end{document}